\documentclass[prodmode,acmcsur,oneside,nonacm]{acmart}	
\usepackage{rotating, graphicx}
\usepackage{longtable}
\usepackage{subcaption}
\usepackage{booktabs}
\usepackage{pgf}
\usepackage{tikz}
\usepackage{listings}
\usetikzlibrary{shapes.geometric}
\usetikzlibrary{arrows}
\usetikzlibrary{positioning}
\usetikzlibrary{automata}
\usepackage{forest,adjustbox}
\usepackage{gb4e}
\usepackage{float}
\usepackage{wrapfig}
\usepackage{hyperref}
\usepackage{xcolor}
\usepackage{multirow}

\definecolor{bostonuniversityred}{rgb}{0.8, 0.0, 0.0}

\ExplSyntaxOn
\tl_new:N \l_listings_boxed_options_tl
\keys_define:nn { listings/boxed }
 {
  caption .tl_set:N = \l_listings_boxed_caption_tl,
  shortcaption .tl_set:N = \l_listings_boxed_shortcaption_tl,
  label .tl_set:N = \l_listings_boxed_label_tl,
  unknown .code:n =
          \tl_put_right:NV \l_listings_boxed_options_tl \l_keys_key_tl
          \tl_put_right:Nn \l_listings_boxed_options_tl { = #1 , },
 }
\box_new:N \l_listings_boxed_box

\lstnewenvironment{blstlisting}[1][]
 {
  \keys_set:nn { listings/boxed } { #1 }
  \exp_args:NV \lstset \l_listings_boxed_options_tl
  \hbox_set:Nw \l_listings_boxed_box
 }
 {
  \hbox_set_end:
  \cs_set_eq:cc {c@figure} {c@lstlisting}
  \tl_set_eq:NN \figurename \lstlistingname
  \tl_if_empty:NF \l_listings_boxed_caption_tl
   {
    \tl_if_empty:NTF \l_listings_boxed_shortcaption_tl
     {
      \captionof{figure}{\l_listings_boxed_caption_tl}
     }
     {
      \captionof{figure}[\l_listings_boxed_shortcaption_tl]{\l_listings_boxed_caption_tl}
     }
    \tl_if_empty:NF \l_listings_boxed_label_tl { \label{\l_listings_boxed_label_tl} }
   }
  \leavevmode\box_use:N \l_listings_boxed_box
 }
\ExplSyntaxOff

\acmJournal{CSUR}

\begin{document}

\title{To Migrate or not to Migrate: An Analysis of Operator Migration in Distributed Stream Processing}

\author{Espen Volnes, Thomas Plagemann and Vera Goebel}
\affiliation{
  \institution{University of Oslo, Department of Informatics}
  \streetaddress{P.O box 1080}
  \streetaddress{Gaustadalléen 23 B}
  \city{Oslo}
  \postcode{0316 Oslo}
  \country{Norway}
}
\email{{espenvol,plageman,goebel}@ifi.uio.no}

\renewcommand{\shortauthors}{E. Volnes et al.}

\begin{abstract}
One of the most important issues in data stream processing systems is to use operator migration to handle highly variable workloads in a cost-efficient manner and adapt to the needs at any given time on demand. Operator migration is a complex process that involves changes in the state and stream management of a running query, typically without any loss of data, and with as little disruption to the execution as possible. This survey provides an overview of solutions for operator migration from a historical perspective as well as the perspective of the goal of migration. It introduces a conceptual model of operator migration to establish a unified terminology and classify existing solutions. Existing work in the area is analyzed to separate the mechanism of migration from the decision to migrate the data. In case of the latter, a cost-benefit analysis is emphasized that is important for operator migration but is often only implicitly addressed, or is neglected altogether. A description of the available solutions provides the reader with a good understanding of the design alternatives from an algorithmic viewpoint. We complement this with an empirical study to provide quantitative insights on the impact of different design alternatives on the mechanisms of migration.
\end{abstract}

\maketitle

\section{Introduction}
\label{sec:introduction}
Stream processing has been researched for more than 20 years, and is becoming ubiquitous in domains of application where just-in-time decision-making is essential \cite{Mehmood2020}, like the Internet of Things (IoT), fraud and anomaly detection, smart cities, and autonomic systems. An indicator of the wide use and importance of stream processing is that most cloud vendors offer support for deploying managed stream processing pipelines \cite{fragkoulis2020survey}, and a sign of its future relevance is the estimated economic impact of the IoT industry, estimated to be between \$3.9 trillion and \$11.1 trillion a year by 2025, around 11\% of the global economy \cite{manyika2015unlocking}. 

Stream processing engines (SPE) come in several flavors, are deployed in different environments (i.e., cloud, fog, edge, in-network), and are called data stream management systems, real-time stream analytics, event stream processing, and complex event processing (CEP). The common denominator in all these systems is that data arrive continuously (generally as tuples) from multiple sources, and need to be processed as soon as they arrive (in memory) to enable immediate decision-making. Thus, the response time must be short even in case of large loads.

SPEs take queries as input and compile them into operator graphs. These are directed acyclic graphs (DAGs) that represent the logical execution of a query, which includes the state management of sub-queries and stream management, i.e., the dependencies between them. If these operators are mapped to several physical hosts and form an overlay network, a distributed stream processing system (DSPS) is established. Incoming tuples to an operator are processed in some way, by transforming, filtering, aggregating, or running a user-defined function.

A key requirement for a DSPS is the ability to handle system dynamics, like changes in workload or resource availability, and potentially mobility. The key mechanism for handling such changes through load balancing and elasticity is operator migration. It is used, for example, to react when applications do not get the guaranteed quality of service (QoS), optimize hardware utilization in data centers, react to failures, and to move operators closer to the data sources when they move. Operator migration entails (1) state management to move the state of the operator from an old host to a new host, and (2) stream management to change data stream routing in the overlay network and, potentially, to buffer data tuples during migration. Decisions on when to migrate the data and where to migrate them to are key aspects of operator migration. The potential approaches to state management, stream management, and decision-making as well as their combinations result in a large design space for operator migration algorithms.

The goal of this work is to give the reader a good understanding of existing solutions to operator migration and the effects of the relevant design decisions. To this end, we develop a conceptual model that captures the fundamental components of operator migration, i.e., the components on which all solutions are based. This model should contribute to operator migration algorithms by providing a unified terminology and establishing a taxonomy. Based on this, we review the literature on operator migration.

Operator migration introduces some form of cost, like freeze time during migration or increased resource consumption to move the state of the operator. Thus, keeping these costs low is a core requirement in the design of operator migration algorithms. Furthermore, during decision-making, it is important to balance the costs of migration against its benefits. There is a general awareness of this trade-off, but surprisingly, few studies have explicitly described how costs and benefits are considered in the migration algorithm and decision-making. Therefore, we place particular emphasis on costs and benefits in our analysis of work in the area. This leads us to two research questions that structure our survey of studies on operator migration:
\begin{itemize}
\item Which mechanisms are used to perform migration?
\item How is the migration decision executed?
\end{itemize}

These research questions are generic, and can be applied to all applications that perform operator migration. However, we focus in this survey on streaming applications to categorize solutions for operator migration based on the conceptual migration model. The criteria for inclusion in the search of the literature are that the studies must be related to operator migration, where the cost of migration is significant, and the QoS must be handled or evaluated with regard to migration. Works where migration is just handled peripherally and the QoS is not important are excluded.

In addition to this functional view of operator migration, we perform an empirical study to gain quantitative insights into different operator migration and decision models. This empirical quantification demonstrates the need for a comprehensive migration model beyond the contribution of the literature. We use Apache Flink \cite{carbone2015apache} and Siddhi \cite{suhothayan2011siddhi}, two operator migration algorithms, and apply part of the NEXMark benchmark \cite{tucker2008nexmark} as workload to measure performance and resource consumption. The aim is to illustrate the quantitative effect of different design decisions.

\paragraph{Available surveys}
The available surveys on data stream processing that include operator migration \cite{lakshmanan2008placement, hummer2013elastic, hirzel2014catalog, de2018distributed, to2018survey, roger2019comprehensive, qin2019enactment, bergui2021survey} do not answer the above research questions, and do not give the reader an insight into the quantitative impact of certain design decisions. Although they do investigate the issue of online reconfigurations of SPEs, none of the studied surveys has adequately investigated different types of operator migration algorithms as well as the relationships among the cost of migration, the decision to migrate, and the migration algorithm.
 
Lakshmanan et al. \cite{lakshmanan2008placement} studied reconfigurations in stream processing and defined a migration model that has similarities to the one developed here. They distinguished between solutions based on where the change is made: either in the network, data, or flow graph. Moreover, different triggers for migrations were studied, such as thresholds, constraint violations, and periodic re-evaluations. However, they did not investigate the different varieties of operator migration in any detail. Hummer et al. \cite{hummer2013elastic} focused on challenges and techniques of high-throughput streaming applications. Hirzel et al. \cite{hirzel2014catalog} cataloged different types of stream processing optimizations, and To et al. \cite{to2018survey} studied state management in stream processing systems but did not examine state migration in much detail. Similarly, De et al. \cite{de2018distributed} explored migration in relation to stream processing and edge computing but did not conduct a critical comparison of the different solutions. We suggest that studying solutions using our conceptual migration model makes such a comparison possible, and deem it necessary to fully answer our research questions. R\"{o}ger et al. \cite{roger2019comprehensive} investigated operator migration but focused on elasticity, whereas we investigate operator migration as a whole. Qin et al. \cite{qin2019enactment} defined a taxonomy for different live reconfigurations in SPEs, including operator migration, but performed only a superficial analysis. Bergui et al. \cite{bergui2021survey} surveyed geo-distributed frameworks and discussed several challenges pertaining to geo-distributed data analytics, where operator migration plays only a minor role in some of the solutions.

The main contributions of this work are as follows:
\begin{itemize}
\item We propose a conceptual model of operator migration that provides a unified terminology and leads to a taxonomy of operator migration. Moreover, this model facilitates the development of new operator migration algorithms.
\item We provide a survey of work on operator migration that analyzes not only current stream management and state management solutions, but also emphasizes a cost-benefit analysis of the migration decision.
\item We report an experimental study involving two migration algorithms on Apache Flink and Siddhi to gain an insight into the quantitative aspects of operator migration.
\end{itemize}

The remainder of this survey is structured as follows: In Section \ref{sec:migration-model}, we introduce the conceptual model of operator migration. In Section \ref{sec:migration-rqs}, we analyze the literature according to the above research questions. In Section \ref{sec:experiments}, we present the experimental evaluation of two migration algorithms for Apache Flink and Siddhi. In Section \ref{sec:reflections}, we discuss future directions of research in the area, and Section \ref{sec:conclusion} provides the conclusions of this survey.

\section{A Conceptual Model of Operator Migration}
\label{sec:migration-model}

In this section, we establish a conceptual model of operator migration to capture the basic concepts and elements on which consensus has been achieved in the literature, and form a unified terminology for operator migration. To keep the presentation of the model concise, we largely refrain, in this section, from referring to the original studies and the terminologies that they use. This is extensively done in Section \ref{sec:migration-rqs}. Since operator migration is the means to change the placement of at least one operator in a DSPS, we start with a description of the initial placement problem before analyzing operator migration. This analysis identifies the basic components of operator migration. These are grouped into two major concerns: (1) stream management to stop, buffer, redirect, and start streams; and (2) state management to establish the current state of the operator at the new operator host, which may require moving the state from the old host to the new one, and starting a replica for the operator on the latter before the former finishes. The cost of operator migration plays an important role in the design of migration mechanisms and the decision on whether to migrate. Therefore, we cover in Section \ref{sec:migration-model-cost-model} the common cost parameters, and, in Section \ref{sec:migration-model-migration-decision}, the migration decision. The latter is triggered, for example, by a change in the system load, and comprises the calculation of a new operator placement as well as a comparison of the benefits of the new placement versus the costs of the migration itself.

Table \ref{table:categories-overview} lists the studies considered in this work that form the foundation of the conceptual model. It classifies them according to the environment of their deployment and the goal of migration, which are important factors for the migration decision and placement. The most common deployment environments for distributed stream processing (DSP) are cloud, fog, and edge networks. \textit{Cloud} has been used to classify data center applications that might handle very high throughputs, and can scale the stems both horizontally and vertically to handle variable traffic loads. The concepts of fog and edge are relatively new terms that seem similar, but have some significant differences. \textit{Edge} computing often focuses on offloading heavy tasks from local resource-constrained devices to either a close base station or a data center. With edge computing, heavy tasks, like machine learning-based inference and videogames, can be executed using smartphones and laptops. \textit{Fog} is an extension of the cloud in which the computing tasks of an application are distributed on multiple devices, including end devices, edge resources, and the cloud itself \cite{yi2015survey}. As such, clients may send most information to a server close to them instead of a centralized data center to reduce energy consumption, congestion on the Internet, and response times for clients.

{\footnotesize\tabcolsep=3pt 
\vspace{-1em}
\begin{longtable}{ |l|l|l| }
\hline
Category & Sub-category & Papers \\
\hline \hline
Deployment environment & Cloud & \citep{shah2003flux, xing2005dynamic, hwang2007cooperative, zhou2008toward, hummer2011dynamic, fernandez2013integrating, lei2014robust, martin2014scalable, heinze2014auto, heinze2014latency, rundensteiner2004cape, gedik2014partitioning, madsen2015dynamic, martin2015user, zacheilas2015elastic, cardellini2016elastic, madsen2016enorm, li2016enabling, liu2016runtime, hochreiner2016elastic, buddhika2017online, madsen2017integrative, lombardi2017elastic, wang2017automating, fang2017parallel, mai2018chi, cardellini2018optimal, fang2018distributed, liu2019adaptive, hoffmann2019megaphone, wang2019elasticutor, del2020rhino} \\
\hline
& Fog & \citep{ahmad2005network, papaemmanouil2009supporting, repantis2007alleviating, repantis2008hot, wang2008potential, rizou2010solving, cardellini2018decentralized, hiessl2019optimal, jonathan2020wasp} \\
\hline
& Edge & \citep{zhou2006efficient, brettlecker2011reliable, kakkad2012migrating, ottenwalder2013migcep, chatzimilioudis2013novel, ottenwalder2014mcep, luthra2018tcep} \\
\hline
Migration goal & Load balancing & \citep{shah2003flux, xing2005dynamic, zhou2006efficient, repantis2007alleviating, repantis2008hot, wang2008potential, hummer2011dynamic, gulisano2012streamcloud, gedik2014partitioning, lei2014robust, martin2014scalable, de2018distributed, zacheilas2015elastic, madsen2015dynamic, lohrmann2015elastic, de2016keep, madsen2016enorm, li2016enabling, tziritas2016improving, liu2016runtime, buddhika2017online, de2017proactive, wang2017automating, fang2017parallel, ma2018optimization, fang2018distributed, wang2019elasticutor, del2020rhino} \\
\hline
& Elasticity & \citep{hummer2011dynamic, gulisano2012streamcloud, martin2014scalable, heinze2015fugu, heinze2014latency, zacheilas2015elastic, wu2015chronostream, lohrmann2015elastic, cardellini2016elastic, de2016keep, xu2016stela, li2016enabling, hochreiner2016elastic, madsen2017integrative, lombardi2017elastic, de2017proactive, cardellini2018decentralized, del2020rhino, jonathan2020wasp} \\
\hline
& Fault tolerance  & \citep{hwang2007cooperative, brettlecker2011reliable, wu2015chronostream, martin2015user, wang2019elasticutor, del2020rhino} \\
\hline
& QoS    & \citep{pietzuch2006network, repantis2008hot, zhou2008toward, wang2008potential, papaemmanouil2009supporting, rizou2010solving, kakkad2012migrating, ottenwalder2013migcep, ottenwalder2014mcep, lohrmann2015elastic, tziritas2016improving, madsen2017integrative, lombardi2017elastic, de2017proactive, mai2018chi, luthra2018tcep, cardellini2018optimal, liu2019adaptive, hiessl2019optimal, del2020rhino, jonathan2020wasp} \\
\hline
\caption{Overview of studies on the categories of operator migration }
\label{table:categories-overview}
\vspace{-3em}
\end{longtable}}

\subsection{Initial placement}
\begin{figure}[htbp]
\vspace{-1em}
\centering
\includegraphics[width=0.6\textwidth]{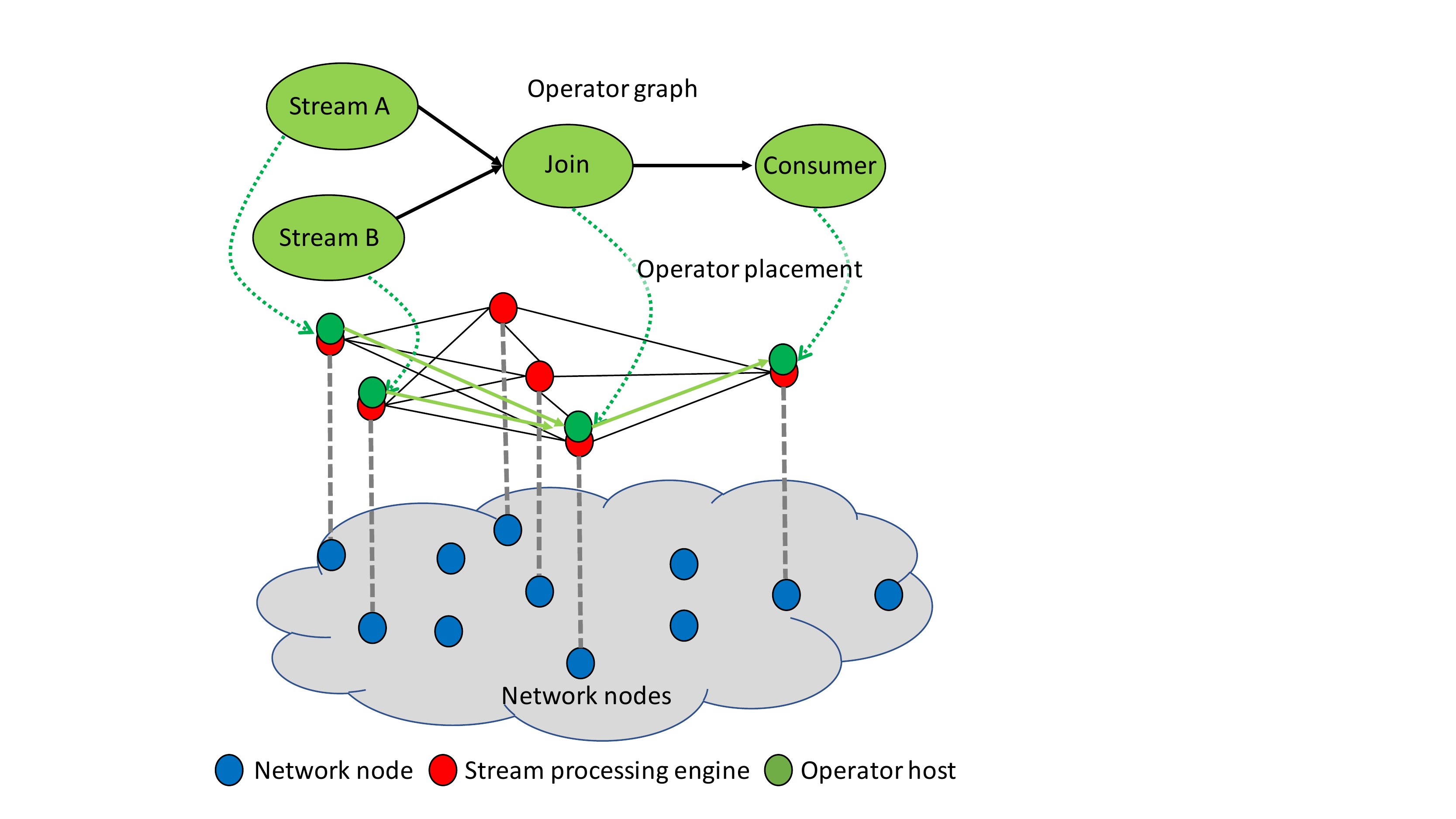}
\caption{Overview of operator placement}
\label{fig:operator-placement}
\vspace{-1em}
\end{figure}

A DSP can be considered to be a set of collaborating SPEs that form an overlay network to process queries over data streams. Consider Figure \ref{fig:operator-placement} as an example of such an application. SPEs run on network nodes that provide the computational and networking resources for the DSP overlay. The objective of the initial operator placement task is to distribute the processing of a query over network nodes such that the goals of the system can be met as adequately as possible. The first step is to transform a query into an operator graph. An operator graph can be modeled as a DAG in which the operators derived from the query are represented as vertices. The placement of these operators in a network, i.e., finding appropriate network nodes to host the operators, is typically driven by an objective function. Such an objective function typically includes (contradicting) criteria of optimization, like low latency of event delivery, low resource consumption (e.g., bandwidth and energy), reliability, and fault tolerance. Typically, a placement function is used to calculate a score for a placement based on the criteria of optimization. To find the optimal placement is an NP-hard problem, and heuristics are often used to find close to optimal solutions. To simplify the discussion, we use the term "optimal" to represent results that are also "close to optimal" based on heuristics. Both centralized and decentralized versions of operator placement can be used to establish an operator network, and are generally implemented as an overlay for the DSP. Nodes in an operator network can be static or mobile, and have one or more of the following roles:
\begin{itemize}
\item \textit{Data producer}: Examples of this include sensors that convert analog signals into data tuples, often with a fixed sampling rate, and software monitors that might create data tuples at a dynamic rate. Obviously, it is important that all produced tuples can be forwarded into the operator network for further processing.
\item \textit{Data consumer}: These are  nodes that request a service, and typically have some QoS requirement, such as tuple latency.
\item \textit{Hosts}: They process at least one operator and contribute to event forwarding in the operator network, i.e., map the input events (from upstream nodes) of the operators they process to output events, and forward them to downstream nodes in the operator network.
\end{itemize}

DSP is typically performed in a dynamic context involving variable workload, resource availability, and mobility. As such, initial placement might, after some time, become sub-optimal and the operator network should be adapted by migrating one or several operators to a new host.

\subsection{Migration mechanism}
The two major concerns of migration mechanisms are state management and stream management. State management is relevant to operators that derive their output based on multiple tuples, e.g., looking for a sequence of tuples using the CEP, joining streams, or aggregating tuples over windows. The \textit{state} can be thought of as tuples. The internal state of the operator is, in practice, typically optimized to include only the necessary information for the given operator, such as the given aggregate value for the extent of a window, or as a finite state machine in the CEP. Therefore, operator migration distinguishes itself markedly from virtual machine (VM) migration, where the entire VM must be transferred to the new host. While some solutions to operator migration, such as the MCEP \cite{ottenwalder2014mcep}, do include VM migration, VM migration is not the focus of this survey. The simplest method of operator migration for a stateful operator is to move it to the new host and replay all necessary historical tuples from the upstream nodes. This technique is used in current publish-subscribe systems, such as Kafka \citep{kafka}, to achieve fault tolerance in stream processing systems like Flink \cite{carbone2015apache}. Using this technique also makes it possible to migrate the data to a different stream processing system, which is usually not possible when extracting the state from the system because the internal state is system specific. However, as the state can grow to become very large, it is often undesirable to replay all tuples. Therefore, this survey focuses on operator migration techniques that extract the state from the stream processing system and move it to the new host.

The task of state management is to establish, at the new host, an operator with the state of the operator at the old host when switching the processing from the old host to the new host. In a \textit{moving state} algorithm, the old host extracts the state of the operator and sends it to the new host. Some algorithms do not need to perform this task, either because they manage stateless operators, e.g., filter operators, or because the old and the new host can schedule a handover. In a \textit{parallel-track} algorithm, originally a term used by Zhu et al. \cite{zhu2004dynamic}, both the old and new hosts receive the same tuples for some time during migration. The handover from the old to the new host is carried out gradually such that the downtime of the operator is minimized. The cost of this approach is that upstream nodes must send twice as many tuples during some part of the migration. A parallel-track algorithm with moving state is called \textit{state-recreation}, and one without moving state is called \textit{window-recreation}. These terms are inspired by StreamCloud \cite{gulisano2012streamcloud}. In a \textit{single-track} algorithm, the upstream nodes send tuples either to the old host or to the new host.

Stream management deals with notifying upstream and downstream nodes of changes made to the topology. Typically, nodes have to update their routing table to reflect the new topology at the upstream node, and this results in a redirection of the outgoing stream to the new operator host. To prevent tuples from getting lost when the operator is down, streams might be stopped and tuples need to be buffered. There are three locations at which tuples can be buffered:  upstream nodes, the old host, and the new host. These tasks of redirecting streams, stopping streams, buffering streams, and restarting streams are coordinated among the hosts involved through control messages. Both centralized and decentralized coordination are possible. As such, there are several design options that can be implemented for a particular operator migration solution.

\begin{figure}[hbtp]
\footnotesize
\begin{forest}
for tree={gray-arrow/.style = {\footnotesize,draw=gray}}
[Migration
    [Migration decision
        [Trigger]
        [Placement]
        [Cost vs. benefit]
        [Proactive vs. reactive]]
    [[Migration algorithm
        [Stream management
            [Stop]
            [Start]
            [Buffer]
            [Redirect]]
        [State management
            [Moving state
                [MoveImmutableState]
                [MoveIncrementalState]
                [MoveState]]
            [Parallel-track
                [ScheduleHandover]]]]]]
\end{forest}
\caption{Concepts of migration }
\label{fig:migration-tree}
\vspace{-1em}
\end{figure}
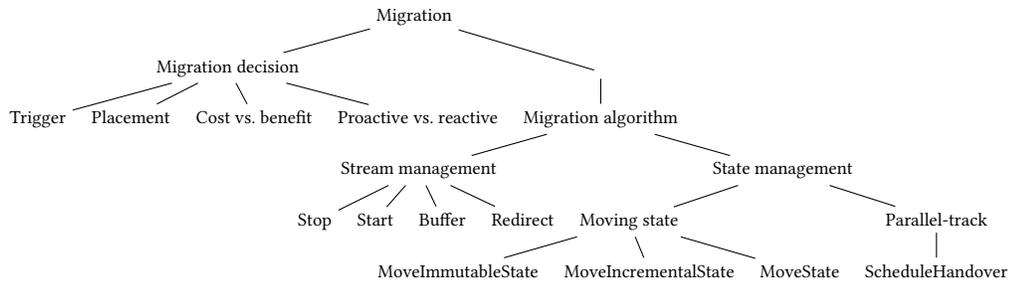

Figure \ref{fig:migration-tree} illustrates the concepts and building blocks that constitute migration algorithms and the relevant decision-making. All algorithms require some stream management functions, such as stop, start, buffer, and redirect. State management differs in that there are many more variations among approaches. Migration decisions first require a trigger for when to make a migration decision and then a placement mechanism to determine the placement that yields the best performance. The cost of the migration must be weighed against its benefit. The degree to which the migration decision should be proactive (e.g., before a host becomes overloaded) or reactive (e.g., when a host is overloaded) must be determined, where the former decisions have higher uncertainty and the latter have a higher cost of migration.

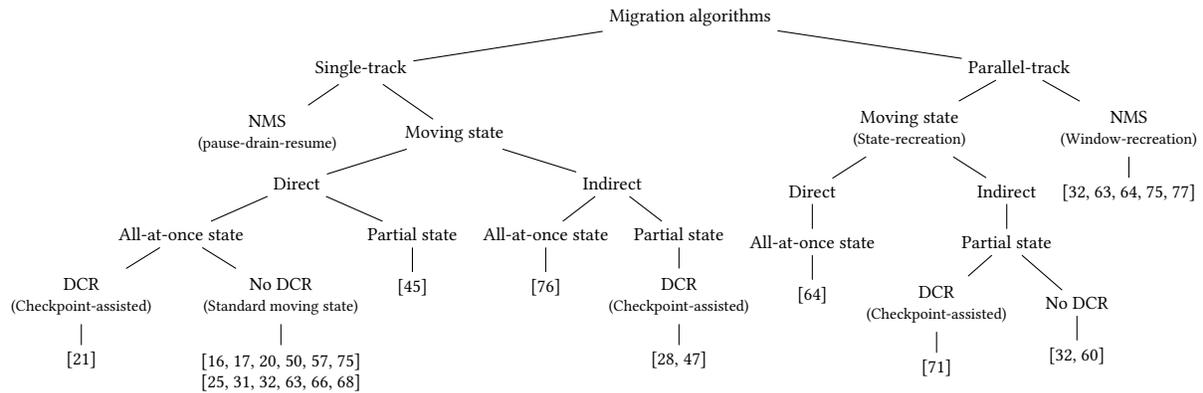
\begin{figure}[hbtp]
\footnotesize
\begin{forest}
[Migration algorithms,
    [Single-track 
        [\begin{tabular}{c}NMS\\\scriptsize{(pause-drain-resume)}\end{tabular}]
        [Moving state, name=moving-single
            [Direct [All-at-once state 
            [\begin{tabular}{c}DCR\\\scriptsize{(Checkpoint-assisted)}\end{tabular} [\citep{fernandez2013integrating}]]
            [\begin{tabular}{c}No DCR\\\scriptsize{(Standard moving state)}\end{tabular} [\begin{tabular}{c}\citep{shah2003flux,zhu2004dynamic,xing2005dynamic,zhou2006efficient,repantis2007alleviating,hummer2011dynamic}\\ \citep{heinze2014latency,madsen2016enorm,cardellini2016elastic,de2016keep,de2017proactive,luthra2018tcep}\end{tabular}]]]
            [Partial state [\citep{hoffmann2019megaphone}]]]
            [Indirect 
            [All-at-once state [\citep{gedik2013elastic}]]
            [Partial state
            [\begin{tabular}{c}DCR\\\scriptsize{(Checkpoint-assisted)}\end{tabular} [\citep{madsen2015dynamic,del2020rhino}]]]]]]
    [Parallel-track, name=parallel
        [\begin{tabular}{c}Moving state\\\scriptsize{(State-recreation)}\end{tabular}, name=moving-parallel
            [Direct [All-at-once state [\citep{gulisano2012streamcloud}]]]
            [Indirect [Partial state 
            [\begin{tabular}{c}DCR\\\scriptsize{(Checkpoint-assisted)}\end{tabular} [\citep{wu2015chronostream}]]
            [No DCR [\citep{ottenwalder2013migcep,madsen2016enorm}]]]]]
        [\begin{tabular}{c}NMS\\\scriptsize{(Window-recreation)}\end{tabular} [\citep{zhu2004dynamic,gulisano2012streamcloud,madsen2016enorm,luthra2018tcep,pham2017uninterruptible}]]]]
\end{forest}
\caption{Migration algorithms}
\label{fig:migration-algorithm-tree}
\end{figure}

Figure \ref{fig:migration-algorithm-tree} shows the different migration algorithms that are formed using the above building blocks. The small text in brackets under some of the categories denotes a term for the given type of algorithm. For instance, a \textit{pause-drain-resume} algorithm is a single-track algorithm without moving state, and, as described earlier, a parallel-track algorithm with moving state is a state-recreation algorithm. The most basic operator migration algorithm is a pause-drain-resume algorithm, and works only with stateless operators or in cases where some state inconsistency is permitted. The operator to migrate is first started on the new host while the old host is also running it. Then, upstream nodes redirect their output streams from the old to the new host. After this, the old host can stop the execution of the operator. Since no state needs to be moved, migration occurs without any downtime. A few control messages must be sent (1) from the controller to the old host, (2) from the old host to the new host, and (3) from the old host to the upstream nodes. As there is no downtime for the operator, any delay caused by these messages is negligible.

When state must be moved and a single-track is used, the operator has some downtime. Specifically, tuples can neither be processed on the old nor the new host while the state is in transmission. In this process, tuples from the upstream nodes must be buffered before they can be processed by the new host. The buffering can be carried out on the upstream nodes, the old host, or the new host. In many cases, tuples can be received after query processing has been stopped. These tuples need to be forwarded from the old host to the new host.

\textit{Partial state} movement involves splitting the state to be migrated into several parts and moving these parts to the new host while the operator is still processing on the old host. This approach avoids having to stop operator processing for the entire state transfer. If the state is periodically checkpointed and distributed on different nodes, this is called \textit{checkpoint-assisted migration}, and can substantially reduce or eliminate the downtime of the operator. Either the entire state already exists on the new host or an incremental checkpoint is extracted before the operator shuts down, and then is sent to the new host. While the last checkpoint is sent to the new host, the operator stops for a much shorter time than when the entire checkpoint is sent at once. A single-track moving state solution can never avoid downtime, because of which parallel-track solutions have been developed.

Parallel-track algorithms differ from single-track algorithms in a fundamental way. They can achieve zero downtime, but at the cost of running the old and new hosts with duplicate input streams and, sometimes, duplicate output streams. A moving-state parallel-track algorithm performs state-recreation, which means that the new host receives the state from the old host while also receiving the same tuples as it does from upstream. A parallel-track algorithm without state migration performs window-recreation, which means that the new host receives the same tuples as the old host until they both have the same tuples in their windows. At this point, the upstream nodes redirect their streams to the new host, and it takes over without the tuples being buffered or any waiting time.

An important motivation for establishing the terminology and building blocks in Figure \ref{fig:migration-algorithm-tree} is that existing work has described the same concepts by different names. For instance, what Zhu et al. \cite{zhu2004dynamic} call parallel-track is described as window-recreation in StreamCloud \cite{gulisano2012streamcloud}, smooth migration in Enorm \cite{madsen2016enorm}, and the seamless minimal state in TCEP \cite{luthra2018tcep}. In StreamCloud, a different algorithm called state-recreation is also parallel-track, but also involves moving state. In contrast to parallel-track, single-track with moving state is called disruptive migration in Enorm \cite{madsen2016enorm} and Pause \& Resume in \cite{heinze2014cloud} because it leads to downtime, as opposed to smooth migration that eliminates downtime. Instant migration is single-track migration without moving state. Checkpoint-assisted algorithms have been described in \cite{wu2015chronostream,madsen2015dynamic,del2020rhino}. Even though these algorithms feature partial state movement, they are different from the Megaphone \cite{hoffmann2019megaphone} as they maintain distributed copies of the state on replicas. As a result, only the minimal state needs to be sent during migration. TCEP has a single-track moving state algorithm called the moving fine-grained state, but it does not use partial state movement as the name suggests. Instead, operators in the operator graph are migrated, one operator at a time, each moving its state all at once.

\subsection{Cost}
\label{sec:migration-model-cost-model}

Operations performed as part of state management and stream management lead to two classes of the cost of migration, related to resource consumption and temporal aspects. The latter are caused by the fact that the operator is not operational during state extraction, state serialization, state movement, state deserialization, and runtime initialization. The bandwidth required to move the state from the old to the new host is the most commonly considered resource in geo-distributed cases. The computational requirements of extracting the state from the old host and the messages needed to coordinate stream management are more commonly considered in centralized data centers. Stream management messages may also have an impact on the duration for which an operator cannot work, e.g., streams from upstream nodes are stopped and no events arrive at the operator until they have been re-directed and started again. The operator is further suspended during state extraction at the old host, moving the state from the old to the new host, and installing it at the new host. 

Two metrics are used to assess temporal cost: \textit{freeze time} and \textit{latency spikes}. Freeze time quantifies the duration for which an operator cannot work, i.e., freeze time = $t_{start}$ - $t_{stop}$, where $t_{stop}$ is the point in time that the old host stops the operator and $t_{start}$ is when the new host resumes it. Latency spikes quantify the increased latency of event delivery caused by a non-working operator. It is often approximated by the time needed for state movement, which is the duration for which the state is in transit between the old and the new host, i.e., state movement time = $t_{receive}$ - $t_{send}$, where $t_{send}$ is the time at which the old host starts sending the state, and $t_{receive}$ is the time at which the entire state has been received at the new host. The state movement time depends on the size of the state and the available bandwidth between old and new host. Thus, state size can be seen as related to the costs of both resources and time. Given these definitions of freeze time and state movement time, it is clear that the freeze time is always longer than the state movement time when state movement requires that processing be paused because the latter does not cover the time for state extraction at the old host, state initialization at the new host, and stream management. Existing research is largely concerned with the tuple delay of a placement \cite{zhou2006efficient}, but tuple latency caused by migration has not been given the same priority.

It should be noted that latency spikes reflect the cost much better than freeze time since it is possible for the operator not to produce any event during the freeze period. Examples of such a case are incoming events during this period that do not match the pattern that can trigger the operator to produce an event, or if a tumbling window implemented by the operator is much larger than the migration time such that all delayed incoming events can be processed on the new host before the window expires.

From the descriptions of the different types of algorithms above, it is easy to see that they differ in the cost of migration. Operator downtime or the latency of the output tuple can be considered a reasonable definition of the cost of migration for single-track state movement algorithms. However, for parallel-track algorithms, this definition of cost can result in excessively frequent migrations because the cost is always close to zero. Therefore, it is necessary to define the cost of migration in such a way that the migrations do not become too frequent.

With parallel-track, operator replicas need to be executed during migration, and upstream nodes must send duplicate streams to the old and new hosts. They may also take a significant amount of time to execute when using window-recreation, which, in addition to using operator replicas during this time, might result in a significant increase in monetary costs. Therefore, it makes sense to consider the monetary cost when using parallel-track algorithms.

\subsection{Migration decision}
\label{sec:migration-model-migration-decision}
Most studies have handled migration as part of an adaptation mechanism, where the goal is to improve execution or recover it in case of node failure. Regularly collected metrics can be used to indicate the need for an adaptation. Recent surveys have focused on adaptation mechanisms \cite{qin2019enactment,fragkoulis2020survey}, and in this work, migration is the main objective of study. Migration is usually the most costly aspect of an adaptation, and this perspective can be useful for better understanding adaptations. Even though adaptations differ in some aspects, they share major parts in terms of cost.

\subsubsection{Migration goal}
This section describes four of the most common goals of migration: \textit{load balancing} to distribute the load evenly on the available nodes, \textit{elasticity} to efficiently leverage computational resources, \textit{fault tolerance} to ensure that the DSPS can continue processing in the event of failures, and improving the QoS. This is not a comprehensive set of migration goals but constitutes categories of fitting for the surveyed papers. While all goals of migration can be applied to any deployment environment, the surveyed solutions with load balancing and elasticity are mainly aimed at cloud-based DSPSs and executed within a single data center, whereas QoS optimization is normally carried out when an operator undergoes backpressure and needs an adaptation to improve the QoS, which can happen in any deployment environment. While migration is relevant for fault tolerance, few solutions describe it as a mechanism to facilitate reliable execution. Instead, solutions often use an upstream rollback approach \cite{koldehofe2013rollback} that replays to the new host tuples that are part of the failed operator. This method is not covered in this survey.

Table \ref{table:migration-reasons-overlap} lists the goals of migration and the overlap between studies in the area in terms of percentage. For instance, 45\% of the surveyed papers on elasticity also consider load balancing. This is a common combination because load balancing can be used after performing a scaling operation to redistribute the load. Few fault tolerance-based solutions describe migration mechanisms, but it is natural that fault tolerance overlaps with load balancing or elasticity as they are often cloud-based solutions, and steps to restore the number of states of a node are similar to those of a scale-in operation. Approaches using QoS constraints on operators to determine when to migrate are often combined with load balancing, as one way to know that the workload must be rebalanced is if the QoS guarantees of an operator have been violated. Below, we analyze all the goals of migration except fault tolerance.

\begin{table}[htbp]
\scriptsize
\begin{tabular}{|l|l|l|l|l|l|}
\hline
Migration goal      & Load balancing & Elasticity & QoS & Fault tolerance \\ \hline\hline
Load balancing        & 100\% (28/28) & 32\% (9/28) & 18\% (5/28) & 7\% (2/28) \\ \hline
Elasticity      & 47\% (9/19) & 100\% (19/19) & 32\% (6/19) & 11\% (2/19) \\ \hline
QoS    & 24\% (5/21) & 29\% (6/21) & 100\% (21/21) & 5\% (1/21) \\ \hline
Fault tolerance & 33\% (2/6) & 33\% (2/6) & 17\% (1/6) & 100\% (6/6) \\ \hline
\end{tabular}
\caption{Goals of migration and the overlap in them in studies in the area }
\label{table:migration-reasons-overlap}
\end{table}
\vspace{-2em}

QoS-driven migration is used to improve the QoS. Typically, the most important QoS parameters that need to be optimized are the bandwidth, availability, and latency. In a mobile setting, the goal of the placement of operators is naturally to remain close to the data producers due to the requirement of low latency, and operators attempt to do so by migrating when a new placement better fulfills goals related to latency and bandwidth than the given one. Another important parameter in mobile settings is energy preservation for resource-constrained nodes. In a cluster setting, the goal is often to ensure that the nodes are not overloaded and that the latency of the tuple is not too high. If the latency of an operator increases significantly, it might be migrated to a node that can provide lower latency.

The basic goal of migration is to improve the QoS. Most solutions are more specific about the goal of migration because finding the optimal solution is usually an NP-hard problem, which is unfeasible to solve for networks of most sizes. A simpler approach is to add constraints to the operator. If the operator cannot fulfill these constraints, it must be relocated. This is typically a much more scalable solution that looks for a placement that is good enough, instead of looking for the optimal solution. It is a push-based manner of letting the coordinator know when the operator needs to be relocated. It should be noted that constraints or thresholds are also often used to achieve the other goals of migration. One characteristic of QoS-based migration is that it is mainly related to the migration of individual operators.

Load balancing is a necessity in distributed streaming systems because many network nodes are available and in use. Because streaming systems often have variable workloads, the coordinator should monitor the resource usage on nodes to ensure that neither the network nor the CPU cores become bottlenecks for the performance of the operators. If resource usage on the nodes is unbalanced, the coordinator moves some of the tasks among the nodes. If these are stateful processes, the tasks to be moved must be paused, moved, and restarted on the new node. Load balancing-driven migration differs from QoS-based migration in the sense that the data consumers do not necessarily benefit much from the balancing, and in that multiple operators are usually migrated through load balancing. However, the decision on when to perform load balancing and where to migrate operators must still take into account the same concerns as for constraint violation, i.e., whether the cost of migration is worth the benefit of the new placement.

Elasticity refers to adding or removing operator replicas that facilitate parallel processing (also called operator scaling). For instance, a query with stateful windows that are grouped by a key can be run in parallel in different threads, where each thread is responsible for a subset of the keys. Four scaling operations are commonly used:
\begin{itemize}
    \item \textit{Scale up}: Create a new process and migrate some partitions of existing threads to it.
    \item \textit{Scale down}: Migrate all partitions of a thread to the other threads and shut it down.
    \item \textit{Scale out}: Create a new worker to which some threads can be moved.
    \item \textit{Scale in}: Remove a worker and move its threads to existing workers.
\end{itemize}

In a cloud setting, the scaling out of a streaming system means adding more servers to a cluster. The streaming system then automatically decides which operators to move to it and potentially scale up. Scaling in means the opposite: A server is removed from the cluster. First, all the server's operators are moved to other servers and some scale-down operations might be performed. Scaling in and out can be modeled as special cases of load balancing. When scaling out, a new container or virtual machine is started on a new machine, which is then added to a load balancing pool. The load balancer can then use this new machine for load balancing. When scaling in, a machine is eliminated from the load balancing pool, and at least its own state must be migrated to the other nodes.

\subsubsection{Starting the migration decision process}
To determine whether migration should be performed, it is necessary to compare the current placement with an alternative placement to estimate the benefits of migration. If these benefits are significantly greater than the costs, migration is beneficial. However, the calculation of a new placement, its benefits, and the related costs might require a non-negligible amount of resources. As such, the naive approach to scheduling a migration decision with a fixed frequency might be too costly. Instead, some form of context awareness needs to be supported to detect changes in the system (e.g., related to workload, resource availability, or mobility) that indicate that there might be a good chance of determining a better placement. The relevance of such changes is generally implied by the goal of migration. Monitoring the runtime system is an important task in the context of detecting such changes. The DSPS can also perform some book-keeping, like the number of operators a node hosts, and trigger a migration decision if a threshold is reached.

Load balancing systems make balancing decisions when the load imbalance of the systems is above a certain threshold. For elasticity-based solutions, checks on whether to automatically scale in or out are similarly performed using thresholds. If the system has a balanced load and its use is still above a given threshold, the system might decide to scale out. If the utilization is below a threshold, the system can scale in. If an operator has latency constraints that are not fulfilled, the coordinator can be notified that migration must occur. The coordinator can either be the node hosting the operator in a decentralized solution or a centralized controller in a data center. In all scenarios, a unit collects metrics from the runtime system in order to make a decision.

\subsubsection{Proactive versus reactive}
Migration decisions can be made reactively or proactively. In the former case, a system migrates when the given situation calls for a change to be made, such as when QoS guarantees for an operator are not fulfilled. Proactive migration decisions rely on predictions about future changes that require migrations.

In several cases, the need for migration scales with its cost. For instance, if the migration is triggered when the tuple rate exceeds a limit and causes QoS violations, more tuples are affected by operator downtime when the need for migration is more pressing. In other words, the more pressing is the need to migrate, the higher is the cost of migration. If a node is over-provisioned, and cannot handle a higher input rate for a given operator, the operator benefits from being migrated to another node. If this situation is detected when the input rate is already too high, a potential migration results in latency spikes for the affected tuples. However, if it is possible to predict that the tuple rate will increase, one can reduce the cost of migration by proactively migrating before the tuple rate becomes too high.

The cost-benefit analysis for making migration decisions is not trivial as the cost of migration is a one-time investment and the benefit from better performance is accumulated over time. When confronted with dynamic surroundings in stream processing scenarios, it makes sense to consider a given placement only for a given amount of time. This time can be regarded as the horizon for which predictions are made. Migration decisions are then made in such a way that the new placement amortizes cost during that time. As such, this time horizon is called amortization time. The notion of working with a limited future horizon for making optimization decisions is also used in model predictive control (MPC), and has been applied by De et al. \cite{de2017proactive} to make proactive scaling decisions.

The more the number of tuples that are impacted, the greater is the extent to which the given option is penalized. However, the number of tuples impacted is an estimate that depends on the accuracy of the prediction. It is possible to assume that tuples are sent evenly across the time window of the horizon, in a single burst, as fast as possible, or a mix between the two. To make such predictions, it is necessary to collect metrics from upstream nodes to determine the density of distribution of the transmitted tuples.

\subsubsection{Cost versus benefit}
Once the decision process has been initiated, it is necessary to determine a better placement and relate its benefits to the costs of the migration to determine whether to migrate. One clear approach to calculating a new placement is to re-run the original placement algorithm with the same objective function. Some of the data needed for calculating a new placement might be available from the monitoring component that triggers the migration decision. In most cases, additional live data must be collected, where this represents a substantial part of the overhead of making the migration decision.

The gain in performance owing to a new placement is generally reflected in the output of the objective function of the old placement versus that of the new placement. By optimizing the objective function during placement, a new placement that delivers the best performance is identified. The problem with simply migrating to the host with the best performance is that the cost of migration might be so high that it is not worth migrating. It might be that a sub-optimal placement is preferred in terms of the objective function owing to a lower migration cost, or maybe that no migration is worth it at all. What makes the comparison of cost and placement performance challenging is that they are not directly comparable. On the one hand, different metrics can be used to determine cost and performance, and on the other, the cost of migration is a one-time investment while placement performance continuously increases the overall benefit as long as there are no changes in the system. As such, there is a need to distinguish between the benefits of placement and migration. The \textit{benefit of placement} simply expresses the difference in placement scores between a new host and the given host, while the \textit{benefit of migration} is calculated based on (1) the cost of migration, (2) the placement performance, and (3) the amortization time.

Three common ways to avoid excessively frequent migrations have been discussed by Lakshmanan et al. \cite{lakshmanan2008placement}: (1) A threshold to ensure that the score of the new placement is significantly better than that of the current placement. (2) If the QoS guarantees of an operator are violated, it triggers a migration, which means that migrations are performed only when necessary. (3) Periodic re-evaluation of the objective function where the interval is set to be reasonably high. In a more recent example, Buddhika et al. \cite{buddhika2017online} regularly calculated interference scores of operators that describe the need for migration, and migrated them to a node where they were subjected to less interference. However, neither Lakshmanan et al. \cite{lakshmanan2008placement} nor Buddhika et al. \cite{buddhika2017online} performed an explicit cost-benefit analysis. This is of interest to us not only to avoid excessively frequent migrations, but to understand why migration is worth it in some cases and not in others based on its costs and benefits. If a placement is an improvement over the given placement, we want to be able to state exactly why the migration is worth performing (or not) in a meaningful and understandable way. The amortization of the cost of migration is a simple goal to understand as long as one weighs the one-time cost of migration against the benefit of the continuous performance of the new placement, but this deliberation is often not presented explicitly.

\section{Existing Solutions}
\label{sec:migration-rqs}

We explore existing work on operator migration, and focus on their migration solutions, and the calculation of the costs and benefits of migration. We use the two research questions posed in Section \ref{sec:introduction} to guide the literature search. In general, the investigated solutions assume complete consistency of the state of the operator. This means that after migration, the new host runs the operator with exactly the same state as the old host did before the migration without any loss of data.

\subsection{Which mechanisms are used to perform migration?}
\label{sec:migration-steps-rq}
As the volume and velocity of data have increased with the emergence of big data \cite{abadi2016beckman}, the simple single-track moving state algorithm has become inadequate. Specialized and innovative solutions that provide no downtime and solutions that leverage fault tolerance mechanisms, such as periodically performed back-ups, have been designed. In this section, we explore the state-of-the-art migration algorithms and provide a historical perspective on innovations proposed. 

Migration algorithms are characterized by their state and stream management. This involves executing certain tasks, such as redirecting, buffering, pausing streams, and moving states between nodes. Moreover, it is important to specify whether these tasks can be executed in parallel and where it is most beneficial to execute them. The most important properties identified in Section \ref{sec:migration-model-cost-model} are whether the algorithms require state migration and how this is performed, and whether they are single-track or parallel-track. Most of the investigated migration algorithms can be derived from these properties. For instance, some algorithms are centralized, and rely on a coordinator, such as \citep{shah2003flux,gulisano2012streamcloud,del2020rhino}, whereas others are decentralized and initiate migration on the operator host, e.g., \citep{repantis2007alleviating,ottenwalder2013migcep}. In some cases, multiple dependent migrations are planned and performed in sequence, but the details of managing multiple migrations are not investigated in this survey. Examples of such algorithms include load balancing, where many keys of an operator may be moved to a new location, and when an operator graph is distributed geographically and several operators are migrated, e.g., in TCEP \cite{luthra2018tcep}.

The most basic algorithms are single-track without state migration, single-track with state migration, and parallel-track without state migration, i.e., window-recreation. These were introduced together by Zhu et al. \cite{zhu2004dynamic} and were later applied to the SPE CAPE \citep{rundensteiner2004cape}. The authors discussed the steps of migration and cost models of the different algorithms. They called them moving state, parallel-track, and pause-drain-resume migration algorithms. Using the terminology established in Section \ref{sec:migration-model}, the moving state algorithm is a single-track moving state, the parallel-track algorithm is a parallel-track without state migration, and the pause-drain-resume algorithm is single-track without state migration. The paper by Shah et al. \cite{shah2003flux} forms the basis for load balancing, and presented a means of repartitioning keys in a key-value-partitioned operator state, which is relevant for cluster-based systems. We characterize this algorithm as a single-track moving state algorithm, but in which the operators are already running on the destination node. In contrast to some studies, e.g., by Qin et al. \cite{qin2019enactment}, we do not consider state movements in load balancing and operator migration to be fundamentally different, and posit that only the entities being migrated are different, i.e., keys are moved instead of operators. In load balancing, the entities being migrated are often a set of keys and their associated states, whereas in operator migration, the entities are usually an operator and its associated state.

\subsubsection{Algorithm descriptions}

This section describes the relevant algorithms in a concise and systematic manner. Since details of what happens in migration algorithms are typically omitted from research papers, our descriptions may deviate to some extent from the original implementations of the migration algorithms considered. For the most significant variations of these algorithms, we show how migration is performed using a figure that illustrates the topology of stream processing and the communication between nodes. The following types of node are used in them: old host (OH), new host (NH), upstream nodes (US), and downstream nodes (DS). The US and DS can both represent one or more nodes, but for the sake of simplicity, only one of them is shown in the figures. Each figure is accompanied by an enumerated description of the steps of the relevant algorithm on the right-hand side, and each step is provided in the figure. The contents of the control messages sent are formatted as a list of tasks that must be executed (shown in subsequent listings). 

Control messages used for migration are typically embedded into the data streams. These tell the nodes that a migration will be performed, and might be used for other coordination tasks. Sometimes, this message is sent only to the old or the new host; at other times, it is sent to the old and the new host, or to upstream nodes, old and new hosts, and downstream nodes. There may be many reasons for notifying different nodes about migration, such as updating the view of where key partitions are maintained, and routing streams. We describe only a subset of control messages for each algorithm, and they describe the essential tasks that should be executed to perform stream and state management tasks. In the illustrations, the control messages described are shown in blue whereas the other messages are shown in red. In addition to the visual representation of the topology and messages sent among nodes, the essential tasks to execute during migration are shown in a listing. The first blue control message from the coordinator, which features in step one of each algorithm, is shown in this listing. All other control messages represent subsequent steps in the algorithm that are described in the first blue control message.

\subsubsection{Standard moving state}
The standard moving state algorithm (see Figure \ref{fig:migration-algorithm-tree}) uses direct state movement between the old and the new hosts, and the entire state is sent all at once. Aside from moving the state, migration requires changing the stream routing. Figure \ref{fig:moving-state-shah2003flux} shows the steps involved in the standard moving state algorithm developed by Shah et al. \cite{shah2003flux}. They proposed an operator called flux that can adapt the state partitioning of the pipelines of dataflow using a state movement algorithm. Other state movement algorithms perform mostly the same steps, but may differ in how stream management is performed or the functions of particular nodes. 

\begin{figure}[hbtp]
\centering
\small
{\begin{minipage}[t]{.4\textwidth}
\begin{flushleft}
\begin{tikzpicture}[->,>=stealth',shorten >=1pt,auto,node distance=2.6cm,semithick]
  \tikzstyle{every state}=[fill=red,draw=none,text=white]

  \node[state]         (OH)                     {$OH$};
  \node[state]         (DS) [above right of=OH] {$DS$};
  \node[state]         (US) [below right of=OH] {$US$};
  \node[state]         (NH) [above right of=US] {$NH$};
  \node[state]         (C) [below right of=NH]  {$C$};

  \path (OH) edge node {4} (NH);

  \node at ($(C)!0.35!(NH)$) [isosceles triangle,inner sep=1pt,thick,rounded corners,minimum size=0.2cm,fill=red,shape border rotate=-180,rotate=-45] {1}; 
  \node at ($(C)!0.2!(OH)$) [isosceles triangle,inner sep=1pt,thick,rounded corners,minimum size=0.2cm,fill=blue,text=white,shape border rotate=-180,rotate=-27] {1}; 
  \node (rect) at ($(OH)!0.35!(NH)$) [draw,thick,minimum size=0.1cm,inner sep=0pt,ellipse,fill=blue,text=white] {State};
  \draw[dotted] (OH) edge node {} (DS);
  \draw[dotted] (NH) edge node {} (DS);
  \draw[dotted, draw=red] (US) edge node {3} (OH.south);
  \draw[dotted, draw=green] (US) edge node {6} (NH.south);
  
  \node at ($(OH)!0.35!(US)$) [isosceles triangle,inner sep=1pt,thick,rounded corners,minimum size=0.2cm,fill=blue,text=white,rotate=-45] {\small 2};
  \node at ($(NH)!0.3!(US)$) [isosceles triangle,inner sep=1pt,thick,rounded corners,minimum size=0.2cm,fill=blue,text=white,shape border rotate=-145,rotate=45] {\small 5};
\end{tikzpicture}
\end{flushleft}
\end{minipage}
\begin{minipage}[t]{.44\textwidth}
\begin{flushright}
\mbox{}\newline\mbox{}\newline
\begin{enumerate}
    \item Coordinator forwards control message
    \item Old host pauses upstream
    \item Upstream stops sending to old host
    \item Old host migrates state to new host
    \item New host resumes the upstream
    \item Upstream starts sending to new host
\end{enumerate}
\end{flushright}
\end{minipage}}
\caption{Moving state \cite{shah2003flux}}
\label{fig:moving-state-shah2003flux}
\end{figure}

Our interpretation of the moving state algorithm's blue migration control message from Step 1 in Figure \ref{fig:moving-state-shah2003flux} is described in Listing \ref{lst:single-track-moving-state}. The upstream nodes buffer, stop, and redirect streams from the old host to the new host. Following this, the task of migrating the state from the old host to the new host is issued to the former, after which the streams are resumed. Instead of stopping the upstream nodes, other solutions \citep{gulisano2012streamcloud, del2020rhino} redirect streams from the upstream nodes to the new host. The new host buffers the streams and starts to process them when the state from the old host has been received and installed. Other solutions send the control message to the old host instead of the upstream nodes \cite{repantis2007alleviating}, or even to the new host \cite{hummer2011dynamic}.

\begin{figure}[htbp]
\vspace{-2em}
\begin{center}
\begin{blstlisting}[basicstyle=\scriptsize,caption={Single-track moving state},label={lst:single-track-moving-state}]
ControlMessage(OH
  ControlMessage(Upstream
    BufferStreams(Streams(query))
    StopStreams(Streams(query)) 
    Redirect(Streams(query), OH, NH)
    ControlMessage(OH, MoveState(query, NH))
    Resume(Streams(query))))
\end{blstlisting}
\end{center}
\vspace{-1em}
\end{figure}

\subsubsection{Parallel-track}

There are two types of parallel-track algorithms: state-recreation and window-recreation algorithms. The difference between them is that state-recreation involves state migration and window-recreation does not. Zhu et al. \cite{zhu2004dynamic} introduced the window-recreation parallel-track migration algorithm. Gulisano et al. \cite{gulisano2012streamcloud} presented both a window-recreation and a state-recreation algorithm, and Ottenwalder et al. \cite{ottenwalder2013migcep} performed state-recreation migrations based on changes in mobility. Madsen et al. proposed a direct window-recreation algorithm in Enorm \cite{madsen2016enorm} and a checkpoint-assisted state-recreation algorithm in \cite{madsen2015dynamic}. ChronoStream \cite{wu2015chronostream} performs a checkpoint-assisted state-recreation migration of state slices to provide horizontal elasticity. UniMiCo \cite{pham2017uninterruptible} (uninterruptable migration of continuous queries) is a direct window-recreation algorithm that can handle both time-based and tuple-based window semantics.

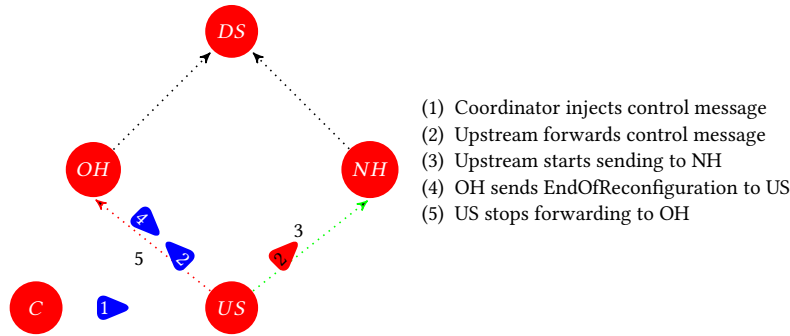
\begin{figure}[hbtp]
\small
{\begin{minipage}[t]{.33\textwidth}
\begin{flushleft}
\begin{tikzpicture}[->,>=stealth',shorten >=1pt,auto,node distance=2.6cm,semithick]
  \tikzstyle{every state}=[fill=red,draw=none,text=white]

  \node[state]         (OH)                     {$OH$};
  \node[state]         (DS) [above right of=OH] {$DS$};
  \node[state]         (US) [below right of=OH] {$US$};
  \node[state]         (NH) [below right of=DS] {$NH$};
  \node[state]         (C)  [left of=US]        {$C$};
  
  \draw[dotted, draw=green] (US) edge node {3} (NH.south);
  \draw[dotted, draw=red] (US) edge node {5} (OH.south);
  \draw[dotted] (OH) edge node {} (DS);
  \draw[dotted] (NH) edge node {} (DS);

  \node at ($(C)!0.35!(US)$) [isosceles triangle,inner sep=1pt,thick,rounded corners,minimum size=0.2cm,fill=blue,text=white] {1};
  \node at ($(US)!0.35!(OH)$) [isosceles triangle,inner sep=1pt,thick,rounded corners,minimum size=0.2cm,fill=blue,text=white,shape border rotate=-180,rotate=-45] {2};
  \node at ($(US)!0.35!(NH)$)[isosceles triangle,inner sep=1pt,thick,rounded corners,minimum size=0.2cm,fill=red,rotate=45] {2};
  \node at ($(OH)!0.35!(US)$) [isosceles triangle,inner sep=1pt,thick,rounded corners,minimum size=0.2cm,fill=blue,text=white,rotate=-45] {\small 4};
\end{tikzpicture}
\end{flushleft}
\end{minipage}
\begin{minipage}[t]{.43\textwidth}
\begin{flushright}
\mbox{}\newline\mbox{}\newline\mbox{}\newline
\begin{enumerate}
    \item Coordinator injects control message
    \item Upstream forwards control message
    \item Upstream starts sending to NH
    \item OH sends EndOfReconfiguration to US
    \item US stops forwarding to OH
\end{enumerate}
\end{flushright}
\end{minipage}}
\caption{Parallel-track window-recreation algorithm \cite{gulisano2012streamcloud}}
\label{fig:parallel-track-window-recreation-streamcloud}
\vspace{-1em}
\end{figure}

\begin{figure}
\small
{\begin{minipage}[t]{.33\textwidth}
\begin{flushleft}
\begin{tikzpicture}[->,>=stealth',shorten >=1pt,auto,node distance=2.6cm,
                    semithick]
  \tikzstyle{every state}=[fill=red,draw=none,text=white]

  \node[state]         (OH)                     {$OH$};
  \node[state]         (DS) [above right of=OH] {$DS$};
  \node[state]         (US) [below right of=OH] {$US$};
  \node[state]         (NH) [below right of=DS] {$NH$};
  \node[state]         (C) [left of=US]         {$C$};

  \draw[dotted, draw=green] (US) edge node {3} (NH.south);
  \draw         (OH) edge node {5} (NH);
  \draw[dotted, draw=red] (US) edge node {6} (OH.south);
  \draw[dotted] (OH) edge node {} (DS);
  \draw[dotted] (NH) edge node {} (DS);

  \node at ($(C)!0.35!(US)$) [isosceles triangle,inner sep=1pt,thick,rounded corners,minimum size=0.2cm,fill=blue,text=white] {1};
  \node at ($(US)!0.35!(OH)$) [isosceles triangle,inner sep=1pt,thick,rounded corners,minimum size=0.2cm,fill=blue,text=white,shape border rotate=-180,rotate=-45] {2};
  \node at ($(US)!0.35!(NH)$) [isosceles triangle,inner sep=1pt,thick,rounded corners,minimum size=0.2cm,fill=blue,text=white,rotate=45] {2};
  \node (rect) at ($(OH)!0.35!(NH)$) [draw,thick,minimum size=0.1cm,inner sep=0pt,ellipse,fill=blue,text=white] {State};
  \node at ($(OH)!0.35!(US)$) [isosceles triangle,inner sep=1pt,thick,rounded corners,minimum size=0.2cm,fill=blue,text=white,rotate=-45] {\small 4};
\end{tikzpicture}
\end{flushleft}
\end{minipage}
\begin{minipage}[t]{.43\textwidth}
\begin{flushright}
\mbox{}\newline\mbox{}\newline
\begin{enumerate}
    \item Coordinator injects control message
    \item Upstream forwards control message
    \item Upstream starts sending to NH
    \item OH sends EndOfReconfiguration to US
    \item OH migrates state to NH
    \item Upstream stops sending to OH
\end{enumerate}
\end{flushright}
\end{minipage}}
\caption{Parallel-track state-recreation algorithm \cite{gulisano2012streamcloud}}
\label{fig:parallel-track-state-recreation-streamcloud}
\end{figure}
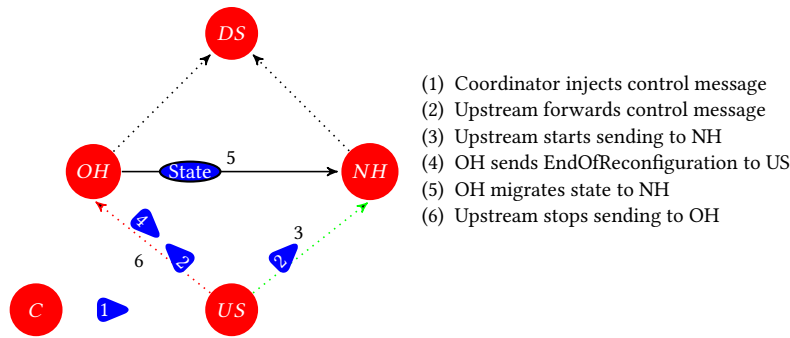

StreamCloud's \cite{gulisano2012streamcloud} state-recreation and window-recreation algorithms are shown in Figure \ref{fig:parallel-track-window-recreation-streamcloud} and Figure \ref{fig:parallel-track-state-recreation-streamcloud}. In both algorithms, a handover between the old and new hosts is scheduled using a timestamp. In window-recreation, the handover is performed in a way such that the old host empties its windows and the new host fills them in parallel, resulting in a smooth handover. For this purpose, the upstream nodes send tuples to both the old and new hosts. In state-recreation, the old host sets the handover timestamp immediately before serializing and transmitting the state to the new host. Any subsequent tuples with a timestamp lower than the handover timestamp are processed by the old host, and the other tuples are processed by the new host. Operator downtime can be avoided here if the handover timestamp is set to a time after the new host is expected to have received the state and started its execution. When the state is received by the new host, it processes all tuples it receives from the upstream nodes in parallel with the old host, but produces only tuples caused by input tuples with a timestamp higher than the handover timestamp.

Our interpretation of the window-recreation algorithm for the blue migration control message from Step 1 in Figure \ref{fig:parallel-track-window-recreation-streamcloud} is described in Listing \ref{lst:parallel-track-window-recreation}. The control message is sent by the coordinator to the upstream nodes. From there, it is forwarded to the old host, which schedules the takeover time for the new host and sends it to the upstream nodes. From then on, the upstream nodes send tuples to both the old and the new hosts. The new host processes the same tuples as the old host but only according to the extent of windows newer than the ones in the old host. It can be assumed that the new host knows that it should not produce any tuple until the extent of the windows have been filled up, i.e., after the old host has stopped processing.

\begin{figure}
\vspace{-2em}
\begin{center}
\begin{blstlisting}[basicstyle=\scriptsize,caption={Window-recreation},label={lst:parallel-track-window-recreation}]
ControlMessage(Upstream
  ControlMessage(NH,
    StartQuery(query))
  ControlMessage(OH
    ControlMessage(Upstream,
      Schedule(RemoveNextHop(Streams(query), OH) 
               TakeoverTime(query))
      AddNextHop(Streams(Upstream), NH))))
\end{blstlisting}
\end{center}
\end{figure}

Our interpretation of the state-recreation algorithm for the blue migration control message from Step 1 in Figure \ref{fig:parallel-track-state-recreation-streamcloud} is described in Listing \ref{lst:parallel-track-state-recreation}. The control message is sent by the coordinator to the upstream nodes. This algorithm requires slightly greater coordination between the old and the new host than in case of window-recreation because the old host must move its state to the new host, and the latter needs to know the takeover time.

\begin{figure}[htbp]
\vspace{-2em}
\begin{center}
\begin{blstlisting}[basicstyle=\scriptsize,caption={State-recreation},label={lst:parallel-track-state-recreation}]
ControlMessage(Upstream 
  ControlMessage(OH
    ControlMessage(NH,
      StopStreams(OutputStreams(query))
      StartQuery(query)
      Schedule(TakeoverTime(query)
               StartStreams(Streams(query))))
    ControlMessage(Upstream,
      Schedule(RemoveNextHop(Streams(query), OH), 
               TakeoverTime(query))
      AddNextHop(Streams(Upstream), NH)))
  MoveState(query, NH))
\end{blstlisting}
\end{center}
\vspace{-1em}
\end{figure}

\subsubsection{Indirect state movement}

Gedik et al. \cite{gedik2013elastic} described a state migration algorithm for load balancing that has been used as the basis in several studies \citep{cardellini2016elastic, li2016enabling, de2017proactive}. They proposed an operator that outputs to multiple replicas partitioned by keys, called a splitter, that can decide to change the distribution of the keys, which requires state migration between replicas. Moreover, they introduced a two-phase approach to migration: donate and collect. In the donate phase, the state to be migrated is moved from the old host's in-memory store to a backing store. In the collect phase, the new host retrieves the state from the backing store. This method was subsequently used by Cardellini et al. \cite{cardellini2016elastic} and Li et al. \cite{li2016enabling} to implement features of elasticity in migration in Apache Storm. The drawback of this method is that streams from the upstream nodes are paused during execution. De et al. \cite{de2016keep, de2017proactive} defined a similar state migration algorithm. However, their implementation contains a number of improvements, e.g., the splitter can send new tuples during state movement instead of blocking until migration is complete. The benefit of the two-phase approach is that it involves an API where an operator simply requires implementing methods to extract the state, and sends it to a backing store instead of requiring intricate communication among operators. Moreover, it can use existing fault tolerance mechanisms that periodically create checkpoints of states for the backing store.

In the donate phase of the algorithm proposed by Gedik et al. \cite{gedik2013elastic}, replicas place the state to be moved into packages, one for each replica that takes over the state. The data are moved away from the in-memory store of the replicas to a backing store. A vertical barrier is used across the replicas to ensure that they do not progress to the next phase until all packages have been donated. In the collect phase, the replicas check the backing store for any packages that contain the state that they take over and restore it. Following this, a horizontal barrier is used to prevent the splitter from sending any tuples until the migration process has been completed.

\subsubsection{Partial state movement}
With partial state movement, the state is partitioned and each partition is moved individually. The aim is to minimize operator downtime in state movement algorithms. MigCEP \cite{ottenwalder2013migcep} is an algorithm designed for frequent migrations to minimize downtime. The state is split up into two parts: immutable and mutable. An immutable or static state includes the operator and, possibly, databases whose data have not changed during migration. A mutable state consists of tuples that are being processed in the operator.

A further improvement involves sending the last checkpoint of the state to the new host before the operator goes down. This is the case in ChronoStream \cite{wu2015chronostream} and Rhino \cite{del2020rhino}, where the state is split into a state before the operator is migrated, and an incremental checkpoint that includes the new state after the first part has been extracted. This can be seen as analogous to the immutable and mutable states described in MigCEP \cite{ottenwalder2013migcep}.

Hoffman et al. \cite{hoffmann2019megaphone} introduced another technique called Megaphone for migrating many keys in an efficient way to minimize latency spikes. In this case, the state is split into many equal-sized parts. Each causes some downtime for the system. However, while the total migration time increases, the spikes due to tuple latency are substantially reduced compared with when the entire state is sent all at once.

Fragkoulis et al. \cite{fragkoulis2020survey} distinguished between all-at-once and continuous state movements, which are identified in this survey as all-at-once and partial state movements, respectively. Megaphone, Rhino, and ChronoStream are characterized in this survey as exemplars of partial state movement, while Fragkoulis et al. categorized Megaphone as using continuous state movement, and Rhino and ChronoStream as using all-at-once state movement. The reason for this difference is that ChronoStream and Rhino rely on \textit{distributed checkpoint replication}, and need to only send the state that has been built up since the last checkpoint. In this survey, migration is further divided by distinguishing between solutions that use distributed checkpoint replication and ones that do not. Megaphone does not use it, and sends the entire state directly from the old host to the new host, whereas ChronoStream and Rhino depend on distributed checkpoint replication. If Rhino and ChronoStream do not use distributed checkpoint replication, this means that the initial checkpoint is sent from the old host to the new host instead of existing on the new host already. Therefore, using partial state movement is not necessarily an indication that multiple states are sent during migration. This is demonstrated in Section \ref{sec:experiments}, where an implementation inspired by Rhino is presented that uses partial state movement without distributed checkpoint replication. When migration starts, the newest checkpoint is sent before an incremental checkpoint.

\subsubsection{Distributed checkpoint replication}
Some solutions leverage fault tolerance mechanisms to improve the scalability and performance of migration using periodically updated, and distributed and replicated checkpoints of the state of stream processing. Since these algorithms use checkpoint solutions that may already exist, they are called checkpoint-assisted algorithms. If the target of migration is a host that already contains the state, a migration algorithm can be as simple as one that loads the checkpoint in memory and replays the upstream tuples to the new host. This requires exactly-once guarantees, as provided by pub-sub systems such as Kafka \cite{kafka}. A parallel-track algorithm can work similarly, but, instead of stopping the old host before replaying tuples on the new host, runs both the old host and the new host until the latter takes over. In this process, output tuples need to be filtered to remove duplicate tuples. ChronoStream \cite{wu2015chronostream} uses distributed checkpoint replication to implement a parallel-track algorithm, Rhino \cite{del2020rhino} to realize a single-track algorithm, and the proposal of Madsen et al. \cite{madsen2015dynamic} to carry out both. The algorithms often also use partial state movement when updating checkpoint replicas to send as little state as possible.

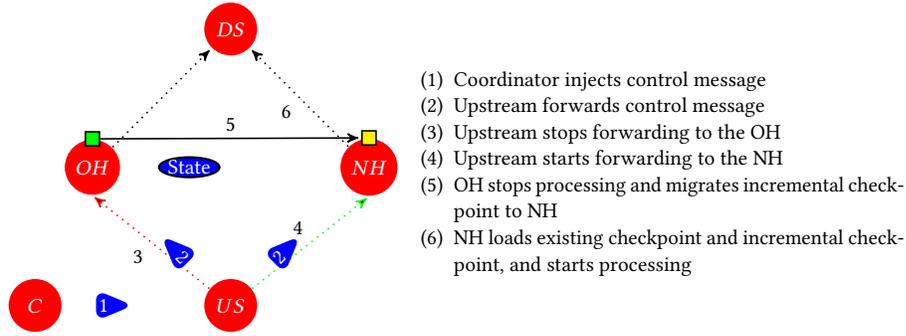
\begin{figure}[hbtp]
\small
{\begin{minipage}[t]{.33\textwidth}
\begin{flushleft}
\begin{tikzpicture}[->,>=stealth',shorten >=1pt,auto,node distance=2.6cm,
                    semithick]
  \tikzstyle{every state}=[fill=red,draw=none,text=white]

  \node[state]         (OH)                     {$OH$};
  \node[draw, fill=green, fit={(OH.north)}] (OH-checkpoint) {};
  \node[state]         (DS) [above right of=OH] {$DS$};
  \node[state]         (US) [below right of=OH] {$US$};
  \node[state]         (NH) [below right of=DS] {$NH$};
  \node[draw, fill=yellow, fit={(NH.north)}] (NH-checkpoint) {};
  \node[state]         (C) [left of=US]         {$C$};

  \draw[dotted] (OH) edge node {}  (DS);
  \draw[dotted] (NH) edge node {6} (DS);
  \draw[dotted, draw=red] (US) edge node {3} (OH.south);
  \draw[dotted, draw=green] (US) edge node {4} (NH.south);

  \path (OH-checkpoint) edge node {5} (NH-checkpoint);

\node at ($(C)!0.35!(US)$) [isosceles triangle,inner sep=1pt,thick,rounded corners,minimum size=0.2cm,fill=blue,text=white] {1};
\node at ($(US)!0.35!(OH)$) [isosceles triangle,inner sep=1pt,thick,rounded corners,minimum size=0.2cm,fill=blue,text=white,shape border rotate=-180,rotate=-45] {2};
\node at ($(US)!0.35!(NH)$) [isosceles triangle,inner sep=1pt,thick,rounded corners,minimum size=0.2cm,fill=blue,text=white,rotate=45] {2};
\node (rect) at ($(OH)!0.35!(NH)$) [draw,thick,minimum size=0.1cm,inner sep=0pt,ellipse,fill=blue,text=white] {State};

\end{tikzpicture}
\end{flushleft}
\end{minipage}
\begin{minipage}[t]{.45\textwidth}
\begin{flushright}
\mbox{}\newline\mbox{}\newline
\begin{enumerate}
    \item Coordinator injects control message
    \item Upstream forwards control message
    \item Upstream stops forwarding to the OH
    \item Upstream starts forwarding to the NH
    \item OH stops processing and migrates incremental checkpoint to NH
    \item NH loads existing checkpoint and incremental checkpoint, and starts processing
\end{enumerate}
\end{flushright}
\end{minipage}}
\caption{Checkpoint-assisted single-track algorithm \cite{del2020rhino}}
\label{fig:checkpoint-assisted-moving-state-rhino}
\end{figure}

Del et al. \cite{del2020rhino} introduce a checkpoint-assisted single-track migration mechanism that can migrate state sizes of up to terabytes 15 times faster than the state-of-the-art solutions by using incremental checkpointing. Their algorithm is shown in Figure \ref{fig:checkpoint-assisted-moving-state-rhino}. Most of the state is sent before the old host is stopped. Afterward, it sends an incremental checkpoint that represents a change in the original state. In this way, only tuples that arrive after migration has started need to be migrated in the incremental checkpoint. This algorithm is a cluster-based migration algorithm that is executed by a handover manager (HM). The HM informs all workers about the migration and about what will happen, by injecting a control message into the source streams, which is a functionality inspired by Chi \cite{mai2018chi}. Afterward, the source nodes are redirected. When the old host has received a control message on all of its incoming streams, it sends the state to the new host. The intermediary hosts, including the old and the new host, send control messages to their next hop nodes. When the nodes have completed their tasks, including the redirection of streams and the migration of state, they acknowledge the HM. The migration is complete when all nodes have acknowledged the HM.

Our interpretation of the checkpoint-assisted moving state algorithm's blue migration control message from Step 1 in Figure \ref{fig:checkpoint-assisted-moving-state-rhino} is described in Listing \ref{lst:moving-state-ca1}. The control message is issued to the upstream nodes, which forward a control message to all downstream nodes. We describe tasks that the old host might be assigned. The main difference between this algorithm and the standard moving state algorithm is that most of the state is assumed to be on the new host before the migration starts. As such, when the state is moved, it is moved using the partial state movement task {\ttfamily{MoveIncrementalState}} instead of {\ttfamily{MoveState}}.

\begin{figure}[htbp]
\begin{center}
\begin{blstlisting}[basicstyle=\scriptsize,caption={Checkpoint-assisted single-track},label={lst:moving-state-ca1}]
# Bootstrapping
ControlMessage(OH, ReplicateCheckpoint(NH))

# Migration
ControlMessage(Upstream 
  ControlMessage(NH, 
    BufferStreams(NH, Streams(query))
    StopStreams(NH, Streams(query)))
  ControlMessage(OH,
    Redirect(Streams(query), OH, NH)
    MoveIncrementalState(query, NH)
    ControlMessage(NH, StartStreams(Streams(query)))))
\end{blstlisting}
\end{center}
\end{figure}

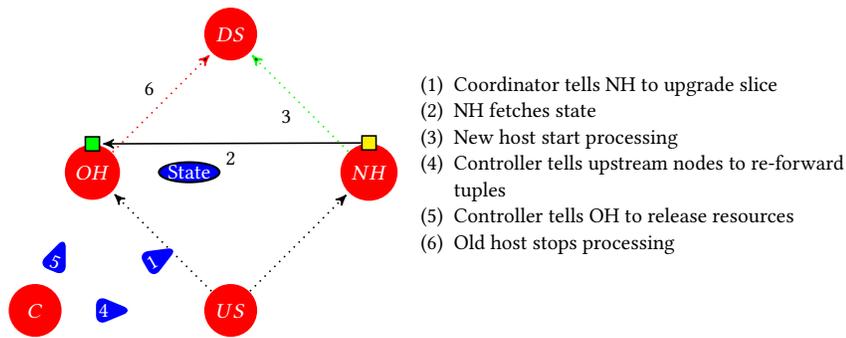
\begin{figure}[hbtp]
\small
{\begin{minipage}[t]{.33\textwidth}
\begin{flushleft}
\begin{tikzpicture}[->,>=stealth',shorten >=1pt,auto,node distance=2.6cm,semithick]
  \tikzstyle{every state}=[fill=red,draw=none,text=white]

  \node[state]         (OH)                     {$OH$};
  \node[state]         (DS) [above right of=OH] {$DS$};
  \node[state]         (US) [below right of=OH] {$US$};
  \node[state]         (NH) [below right of=DS] {$NH$};
  \node[state]         (C)  [left of=US]        {$C$};
  
  \node[draw, fill=green, fit={(OH.north)}] (OH-checkpoint) {};
  \node[draw, fill=yellow, fit={(NH.north)}] (NH-checkpoint) {};

  \path (NH-checkpoint) edge node {2} (OH-checkpoint);

  \node at ($(OH)!0.35!(NH)$) [draw,thick,minimum size=0.1cm,inner sep=0pt,ellipse,fill=blue,text=white] {State};
  
  \draw[dotted, draw=green] (NH) edge node {3} (DS);
  \draw[dotted, draw=red] (OH) edge node {6} (DS);
  \draw[dotted] (US) edge node {} (OH);
  \draw[dotted] (US) edge node {} (NH);
  \node at ($(C)!0.35!(NH)$) [isosceles triangle,inner sep=1pt,thick,rounded corners,minimum size=0.2cm,fill=blue,text=white,rotate=30] {\small 1};
  \node at ($(C)!0.35!(OH)$) [isosceles triangle,inner sep=1pt,thick,rounded corners,minimum size=0.2cm,fill=blue,text=white,shape border rotate=90,rotate=-22.5] {\small 5};
  \node at ($(C)!0.35!(US)$) [isosceles triangle,inner sep=1pt,thick,rounded corners,minimum size=0.2cm,fill=blue,text=white] {\small 4};

\end{tikzpicture}
\end{flushleft}
\end{minipage}
\begin{minipage}[t]{.40\textwidth}
\begin{flushright}
\mbox{}\newline\mbox{}\newline
\begin{enumerate}
    \item Coordinator tells NH to upgrade slice
    \item NH fetches state
    \item New host start processing
    \item Controller tells upstream nodes to re-forward tuples
    \item Controller tells OH to release resources
    \item Old host stops processing
\end{enumerate}
\end{flushright}
\end{minipage}}
\caption{Checkpoint-assisted parallel-track algorithm \cite{wu2015chronostream}}
\label{fig:checkpoint-assisted-parallel-track-chronostream}
\end{figure}

Wu et al. proposed ChronoStream \cite{wu2015chronostream}, a checkpoint-assisted state-recreation migration algorithm that provides horizontal elasticity, as illustrated in Figure \ref{fig:checkpoint-assisted-parallel-track-chronostream}. The states of all tasks on a node are periodically backed up and sent to the other nodes. As a result, migration only involves updating a subset of the backed-up state, which significantly reduces the number of states to be moved. This process is split into four phases: migration preparation, state rebuilding, dataflow rerouting, and resource release. The first phase sets up a container for the operator on the destination node if this has not been already done. In the second phase, the new host fetches the operator’s state locally or remotely and rebuilds it, and notifies the master node when finished. The third phase involves the master telling the data sources to send tuples to the new host as well, including any tuple that is not included in the state that the new host received. At this point, the new host participates in the processing and produces the same tuples as the old host, and duplicate output tuples are filtered out by downstream operators based on the sequence numbers of the tuples. Finally, the controller tells the old host to release the resources such that the new host is the only node running the operator.

Our interpretation of the checkpoint-assisted parallel-track algorithm for the blue migration control message in Figure \ref{fig:checkpoint-assisted-parallel-track-chronostream} is described in Listing \ref{lst:parallel-track-ca}. The main difference between the parallel-track checkpoint-assisted algorithm and a non-checkpoint-assisted algorithm is that the immutable state is sent before anything else happens.

\begin{figure}[htbp]
\vspace{-2em}
\begin{center}
\begin{blstlisting}[basicstyle=\scriptsize,caption={Checkpoint-assisted parallel-track},label={lst:parallel-track-ca}]
# Bootstrapping
ControlMessage(OH, ReplicateCheckpoint(NH))

# Migration
ControlMessage(US, AddNextHop(Streams(query), NH)
  RemoveNextHop(Streams(query), OH))
ControlMessage(NH, 
  ControlMessage(OH, MoveImmutableState(query, NH)))
ControlMessage(OH, StopQuery(query))
\end{blstlisting}
\end{center}
\vspace{-2em}
\end{figure}

\subsection{How is the migration decision executed?}
\label{sec:migration-decisions-rq}

We first provide an overview of the parameters of optimization, and the cost and benefit metrics used in existing work (Table \ref{table:optimization-goals} and Table \ref{table:migration-cost-RQ}). We then describe (1) how cost values are estimated and measured, (2) approaches for optimization to increase benefits, and (3) reactive and proactive methods.

Even though there are many different definitions of the parameters of optimization, they are often related. Therefore, we group them in Table \ref{table:optimization-goals} into five categories: network performance (e.g., bandwidth, bandwidth latency product), tuple performance (e.g., tuple latency, tuple rate), load, costs of migration, and monetary costs. Since the goal of migration is important for optimization, we differentiate among the categories of parameters of optimization in research according to the goals of migration. The most prominent goal is load balancing, and load is the most commonly used optimization parameter. While monetary cost is the least commonly used optimization parameter, migration is often used to avoid the need for over-provisioning, and thus, indirectly saves money.

\begin{table}[htbp]
\footnotesize
\begin{tabular}{|l|l|l|}
\hline
Parameter               & Migration goal & Papers \\ \hline\hline
Tuple performance       & Load balancing     & \citep{hummer2011dynamic,liu2016runtime,buddhika2017online,wang2008potential,lohrmann2015elastic,lohrmann2015elastic} \\ \hline
                        & Elasticity   & \citep{hummer2011dynamic,heinze2014latency,lohrmann2015elastic,xu2016stela,de2017proactive} \\ \hline
                        & QoS      & \citep{lohrmann2015elastic, jonathan2020wasp} \\ \hline
Network performance     & Load balancing & \citep{wang2008potential, lei2014robust} \\ \hline
                        & Elasticity & \citep{zacheilas2015elastic} \\ \hline
                        & QoS    & \citep{pietzuch2006network, wang2008potential, zhou2008toward, rizou2010solving, kakkad2012migrating, ottenwalder2013migcep, ottenwalder2014mcep, cardellini2018optimal, hiessl2019optimal} \\ \hline
Load                    & Load balancing   & \citep{xing2005dynamic, zhou2006efficient, repantis2008hot, wang2008potential, hummer2011dynamic, gedik2014partitioning, lei2014robust, lohrmann2015elastic, liu2016runtime, buddhika2017online, fang2018distributed, wang2019elasticutor} \\ \hline
                        & Elasticity & \citep{hummer2011dynamic, lohrmann2015elastic, zacheilas2015elastic, hochreiner2016elastic, de2017proactive, lombardi2017elastic, cardellini2018decentralized} \\ \hline
                        & QoS    & \citep{pietzuch2006network, repantis2008hot, zhou2008toward, kakkad2012migrating, lohrmann2015elastic, lombardi2017elastic, hiessl2019optimal} \\ \hline
                        & Fault tolerance  & \citep{hwang2007cooperative} \\ \hline
Migration costs         & Load balancing   & \citep{repantis2008hot, lei2014robust, gedik2014partitioning, fang2018distributed, wang2019elasticutor} \\ \hline
                        & Elasticity & \citep{zacheilas2015elastic, de2017proactive, cardellini2018decentralized} \\ \hline
                        & QoS    & \citep{repantis2008hot, kakkad2012migrating} \\ \hline
Monetary costs          & Elasticity & \citep{zacheilas2015elastic, hochreiner2016elastic, cardellini2018decentralized} \\ \hline
                        & QoS    & \citep{cardellini2018optimal, hiessl2019optimal} \\ \hline
\end{tabular}
\caption{Goals of optimization grouped by the goal of migration}
\label{table:optimization-goals}
\vspace{-2em}
\end{table}

Table \ref{table:migration-cost-RQ} gives an overview of the metrics used to define the estimated and measured costs of migration, the estimated placement, and the measured benefit of migration. Ideally, the measured and estimated metrics should be identical but they are not. One reason for this mismatch is that it is much easier to measure values for certain metrics than to predict them, like tuple latency and tuple rate. Values for costs and benefits need to be estimated for each decision to migrate before the migration is performed, whereas values of the evaluation metrics are measured during migration. The mismatch between the estimated cost and the benefit, and the measured evaluation metrics might also help complement future migration decisions and the assessment of migration using further metrics. The most commonly used parameters to determine the cost of migration are the migration time and state size, and few systems use more precise cost parameters, such as latency spike and performance penalty. While the columns in the table below may indicate the goals of optimization used, it is different from Table \ref{table:optimization-goals} in that it looks only at what metrics are used. Approaches like \cite{lei2014robust} are listed as those for placement optimization based on the cost of migration, but are not located anywhere in the modeled cost of migration column as no metric is used to describe the cost of migration in the paper.

\begin{table}[hbtp]
\footnotesize
\centering
    \begin{tabular}{|c|c|p{2.5cm}|p{2.5cm}|p{2.5cm}|p{2.5cm}|}
    \hline
    & Metric & Modeled migration cost & Measured migration cost & Modeled placement benefit & Measured benefit of migration \\
    \hline \hline
    \multirow{4}{*}{\rotatebox[origin=c]{90}{Network}} & Link bandwidth & & & \citep{jonathan2020wasp} & \\
    \cline{2-6}
    & Network usage         & & & \citep{wang2008potential} & \citep{pietzuch2006network,wang2008potential}\\
    \cline{2-6}
    & Bandwidth delay product & \citep{ottenwalder2013migcep, ottenwalder2014mcep} & & & \\
    \cline{2-6}
    & Stream duplication & & & \citep{hummer2011dynamic} & \\
    \hline
    \multirow{2}{*}{\rotatebox[origin=c]{90}{\scriptsize{Tuple perf.}}} & Tuple latency & \citep{heinze2014latency} & \citep{shah2003flux, heinze2014auto, heinze2014latency, madsen2015dynamic, madsen2016enorm, de2017proactive, cardellini2018optimal, cardellini2018decentralized, wang2019elasticutor, del2020rhino} & \citep{repantis2008hot,lohrmann2015elastic,zacheilas2015elastic,liu2016runtime,buddhika2017online,cardellini2018optimal,jonathan2020wasp} & \citep{shah2003flux,zhou2006efficient,pietzuch2006network,repantis2008hot,wu2015chronostream,cardellini2016elastic,liu2016runtime,buddhika2017online,lombardi2017elastic,luthra2018tcep,cardellini2017optimal,fang2018distributed,cardellini2018decentralized,liu2019adaptive,hiessl2019optimal,hoffmann2019megaphone,wang2019elasticutor,del2020rhino,jonathan2020wasp} \\
    \cline{2-6}
    & Tuple rate & & \citep{shah2003flux, madsen2016enorm, lombardi2017elastic, wang2019elasticutor, del2020rhino} & \citep{repantis2008hot,hummer2011dynamic,heinze2014latency,xu2016stela,liu2016runtime,lombardi2017elastic,de2017proactive,buddhika2017online,jonathan2020wasp} & \citep{shah2003flux,repantis2008hot,gedik2013elastic,lohrmann2015elastic,wu2015chronostream,buddhika2017online,lombardi2017elastic,fang2017parallel,cardellini2017optimal,fang2018distributed,cardellini2018decentralized,liu2019adaptive,wang2019elasticutor} \\
    \hline
    \multirow{4}{*}{\rotatebox[origin=c]{90}{Load}} & Resource usage & \citep{ma2018optimization} & \citep{heinze2014latency, de2017proactive, hoffmann2019megaphone, del2020rhino} & \citep{xing2005dynamic,repantis2008hot,wang2008potential,hummer2011dynamic,gedik2014partitioning,lei2014robust,lohrmann2015elastic,liu2016runtime,buddhika2017online,fang2018distributed,wang2019elasticutor,zacheilas2015elastic,hochreiner2016elastic,lombardi2017elastic,cardellini2018decentralized,zhou2008toward,hiessl2019optimal,pietzuch2006network,kakkad2012migrating,hwang2007cooperative} & \citep{wang2008potential,gedik2013elastic,cardellini2016elastic,buddhika2017online,lombardi2017elastic,liu2019adaptive,hiessl2019optimal,hoffmann2019megaphone,del2020rhino} \\
    \cline{2-6}
    & Availability          & & & & \citep{hiessl2019optimal}\\
    \cline{2-6}
    & Load shedding             & & & & \citep{liu2019adaptive,jonathan2020wasp} \\
    \hline
    \multirow{7}{*}{\rotatebox[origin=c]{90}{Migration-exclusive}} & \# Migrations         & & \citep{pietzuch2006network, repantis2007alleviating, repantis2008hot, rizou2010solving, tziritas2016improving, buddhika2017online, de2017proactive} & & \\
    \cline{2-6}
    & \# Migration messages & & \citep{repantis2007alleviating, rizou2010solving, tziritas2016improving} & & \\
    \cline{2-6}
    & Number of control messages & \citep{luthra2018tcep} & & & \\
    \cline{2-6}
    & State size & \citep{hummer2011dynamic, gedik2014partitioning, liu2016runtime, madsen2017integrative, wang2017automating, fang2017parallel, fang2018distributed, wang2019elasticutor} & \citep{zhou2006efficient, gulisano2012streamcloud, gedik2014partitioning, cardellini2016elastic, luthra2018tcep, fang2018distributed} & & \\
    \cline{2-6}
    & Migration time & \citep{zhu2004dynamic, xing2005dynamic, zacheilas2015elastic, madsen2015dynamic, martin2015user, madsen2016enorm, lombardi2017elastic, cardellini2018optimal, luthra2018tcep, ma2018optimization, jonathan2020wasp} & \citep{rundensteiner2004cape, hwang2007cooperative, brettlecker2011reliable, hummer2011dynamic, gulisano2012streamcloud, fernandez2013integrating, martin2015user, zacheilas2015elastic, madsen2015dynamic, wu2015chronostream, cardellini2016elastic, xu2016stela, madsen2016enorm, wang2017automating, luthra2018tcep, cardellini2018optimal, hoffmann2019megaphone, wang2019elasticutor, jonathan2020wasp} & & \\
    \cline{2-6}
    & Stabilization time          & & \citep{brettlecker2011reliable, xu2016stela, jonathan2020wasp} & & \\
    \hline
    \end{tabular}
\caption{How the costs of migration are modeled and measured, and its benefits are measured}
\label{table:migration-cost-RQ}
\vspace{-4em}
\end{table}

\subsubsection{Migration cost}

Accurately defining the cost of migration is essential for making the correct migration decisions. The column for the modeled cost of migration in Table \ref{table:migration-cost-RQ} describes the metrics used to represent cost in existing solutions. The column of the measured cost of migration to the right describes the metrics used in the evaluation of existing solutions. It can be manifested in any kind of a degradation of execution, such as decreased throughput or increased tuple latency. The table shows that the metrics used to measure the cost of migration are not the same as the ones used to model it.

The vast majority of solutions use migration-specific metrics to model its cost, as opposed to metrics that are used for measuring its cost and benefit. Tuple processing performance is used in many cases to model and measure benefit, but very few approaches have used it to calculate the cost of migration. Heinze et al. \cite{heinze2014latency} modeled and predicted tuple latency as part of the cost of migration but provided no solution to model the tuple rate because it is much easier to measure tuple processing performance than to predict it. Operator downtime is an indicator of spikes in latency but also depends on the tuple rate.

The migration time and the size of the state to be moved are the most common costs of migration metrics. Migration time is typically calculated as a function of state size, bandwidth, and latency. In environments where the bandwidth and latency are stable, such as within data centers, the state size is often interchangeable with migration time. In most cases, the migration time is assumed to be easy to model and no calculation for it is given. Some solutions can migrate multiple operators at a time, and thus define the migration time as the maximum time it takes to move any of the operators \cite{cardellini2018optimal,jonathan2020wasp}. Cardellini et al. \cite{cardellini2018optimal} used a data center-based solution to define the operator downtime based on the type of adaptation made, size of the state to be moved, and the round-trip delay between the nodes and the computational resources. WASP \cite{jonathan2020wasp} is a wide area network solution that defines the time it takes to move an operator based on the state size and bandwidth between links, the latter of which is significantly more limited and variable in a wide area network than a data center. Zhu et al. \cite{zhu2004dynamic} focused on the time needed for each step of migration, such as the time spent cleaning the accumulated tuples, state matching, moving the state, and recomputing it.

Using state size as the cost of migration is among the easiest ways of defining cost because it requires only looking at the size of the state to be migrated. The solutions that we survey that use state size as cost metric are all cloud based, which makes sense since data centers feature a high and stable bandwidth between nodes, in contrast to geo-distributed environments. The state size is frequently used as part of the objective function when making migration decisions \cite{hummer2011dynamic,gedik2014partitioning} as part of a constraint to prevent costly solutions from being selected \cite{madsen2017integrative}, and can even be the only criterion to minimize when making load balancing decisions \cite{fang2017parallel,fang2018distributed}.

Luthra et al. \cite{luthra2018tcep} used the number of control messages during migration as part of the definition of the cost of migration. This parameter is significant because if nodes have to wait for acknowledgments for these messages, the total migration time then depends on the distance between nodes. When the cost of migration is defined in terms of migration time, only the time taken to move the state is generally included in the equation, and might result in an inaccurate view of the cost.

The bandwidth delay product is a measure of how much data can be sent in a given duration. As part of the cost of migration, it represents the amount of data that can be sent when a migration is underway. The more tuples that can be sent, the higher is the cost of migration, and the less desirable a migration is. MigCEP \cite{ottenwalder2013migcep,ottenwalder2014mcep} uses the average bandwidth delay product during migration as its cost. This represents the utilization of the network due to migration.

\subsubsection{Benefit}
In this section, we discuss the goals of optimization of different migration solutions. This includes the most important metrics used for determining the benefit of migration. The benefit of migration is based on performance in terms of the placement, amortization time, and the cost of migration (as explained in Section \ref{sec:migration-model-cost-model}). This is either explicitly defined or implicit in the decision-making, where the goal is to maximize performance in terms of the placement and minimize the cost of migration.

One could argue that all goals of optimization are relevant to all goals of migration. However, some are more tightly coupled than others. For instance, load balancing involves using the load of a system to make balancing decisions. QoS solutions, on the contrary, are not bound specifically to any goal of optimization. Elasticity-based solutions aim to minimize resource usage while maintaining the QoS. In other words, they use as few resources as possible for an application, and trigger a scaling operation when the load is above or below a given threshold. Fault tolerance-based solutions involve migrations when nodes fail and the operators must be migrated to new or existing nodes.

Network performance as a goal of optimization means using the quality of the network links to determine performance in terms of placement. Important metrics in this context include the bandwidth between links in the overlay topology, the latency between nodes, and the bandwidth delay product. Tuple processing performance in query processing is the most popular indicator of the quality of an adaptation, as shown by the number of studies that have measured the benefit of migration in terms of tuple latency or rate. If a node is overloaded in a data center, the latency of the tuple might exceed acceptable levels, leading to QoS violations. A long migration time might temporarily worsen performance, but if the general gain in performance outweighs the degradation in it, the migration is considered worth it. The load of a system is an important goal of optimization that makes it possible to run as many operators on a node as it can handle, and to make changes when the workload is above or below a given threshold. The cost of migration is essential to consider when making migration decisions to avoid excessively frequent migrations and ensure that the benefit of the new placement outweighs the cost of migration. When the cost of migration is used to calculate its benefit, the result is the modeled benefit of migration. Monetary cost can be useful as a goal of optimization to make a tradeoff between the cost of resources and the performance of the system.

\paragraph{Network}
In decentralized fog and edge computing solutions, network usage as well as bandwidth and latency between links are crucial metrics. Pietzuch et al. \cite{pietzuch2006network} developed an overlay network that can make network-aware placement and migration decisions. Parameters, like the latency and bandwidth of overlay links and the load on nodes are used as criteria of optimization when placing and migrating operators. Rizou et al. \cite{rizou2010solving} implemented a similar method that converges to the optimal placement in fewer migrations than Pietzuch's solution.

\paragraph{Tuple performance}
In a resource-constrained environment, tuple latency can be an indicator of the energy consumption and the goal to minimize latency can implicitly lead to energy reduction. For most surveyed approaches, the goal of migration directly or indirectly involves improving performance. Most elasticity-based and load balancing-based solutions are cluster-based, and are more concerned with the load on the system than the bandwidth of or latency between links. The tuple rate of a data stream is an indicator of the load on the nodes, and can be used to calculate the variance in load. For instance, Buddhika et al. \cite{buddhika2017online} proposed a methodology to reduce the interference between stream processing operators using migration. To achieve this, the interference score of an operator is calculated, where the higher the score is, the greater the need is for migration. This interference score is based on the prediction of future packet load. Repantis et al. \cite{repantis2008hot} defined latency constraints on the operators and used tuple latency to determine when an operator must be migrated.

\paragraph{Load}
Unsurprisingly, all load balancing solutions use either load as a parameter when making decisions or tuple performance to estimate load. Gedik et al. \cite{gedik2014partitioning} propose two methods. First method minimizes the variance in load between nodes in a cluster. In this case, a coordinator monitors the load on the systems and, when a balancing decision has to be made, selects the configuration with the least variation in load. The second method triggers a load balance when the imbalance has crossed over a given threshold to re-balance the load to at least below another threshold. In other words, load balancing is used as a constraint. In this case, the goal is to minimize the cost of migration by redistributing the minimum amount of load to achieve an acceptable load balance. The benefit of the latter over the former method is that the former might require expensive migrations of large loads among many nodes, and redistributing loads that are not the cause of the imbalance. On the contrary, the latter method achieves an acceptable load balance while moving the smallest load.

Elasticity-based solutions increase or reduce the number of resources used by an application based on its variable workload. If a cluster is overloaded after load balancing, this is a sign that the system should scale out \cite{lohrmann2015elastic}. In decentralized fog-based solutions, the load of a system is not known beforehand, and therefore, there might be a tradeoff between latency and load. Pietzuch et al. \cite{pietzuch2006network} introduced a cost space model in which a topology of systems is constructed based on the latency and bandwidth between nodes as well as the load on systems. If the load of a system is large, the relevant node appears farther in the cost space when mapping an operator graph to a physical topology, and thus is less likely to be selected.

\paragraph{Monetary costs}
In the cloud-based model followed by fog and edge models, users mostly pay based on usage. Users can allocate a certain amount of resources and scale out or in whenever more or less resources are needed, respectively, and this is paid for based on usage. A complicated issue in this case is balancing the monetary costs with the benefits of improved placement. None of the load balancing solutions uses monetary cost as an optimization criterion. This makes sense as the load balancing problem involves evenly distributing the load over a fixed amount of resources, whereas elasticity can increase or reduce the amount of resources. In terms of hardware resources, there is nothing to optimize as they are already paid for. The monetary cost of moving states during migration can thus be minimized. Typically, this is implicitly done by designing the objective function to minimize the number of state that need to be moved. Elasticity-based solutions require a tradeoff between resource usage and monetary costs \cite{zacheilas2015elastic, hochreiner2016elastic, cardellini2018decentralized}. An elastic solution might use a threshold for the load to determine when to scale out. However, deciding when to scale in might be more complex, considering that it requires a certain downtime for the worker to be removed.

\paragraph{Costs of migration}
Most studies prevent the cost of migration from affecting the QoS by implicitly minimizing the number of migrations, their frequency, or their magnitude. Zhou et al. \cite{zhou2008toward} emphasized the need to minimize the time needed for query migration but did not describe a means of implementing this in their solution. Lombardi et al. \cite{lombardi2017elastic} defined the cost of migration in terms of the time it takes to perform different steps but did not attempt to minimize it. The cost of migration can be minimized by either using single-objective optimization \citep{gedik2014partitioning, fang2018distributed, wang2019elasticutor, jonathan2020wasp}, or simple additive weighting (SAW) with multiple objectives \citep{hummer2011dynamic, kakkad2012migrating, lei2014robust, cardellini2018optimal, cardellini2018decentralized}. If only the cost of migration is minimized, constraints have to be placed on the quality of the placement to ensure that the selected placement is acceptable. With load balancing, minimizing the cost of migration while maintaining constraints on the load imbalance is a good way to ensure a balanced load that minimally affects the performance of the system.

Minimizing the number of migrations is a similar goal to minimizing the cost of migration. Repantis et al. \cite{repantis2008hot} proposed a hotspot alleviation-based solution with the goal of minimizing the number of migrations that leads to an acceptable QoS for the operators. Rizou et al. \cite{rizou2010solving} implemented a similar relaxation algorithm to the one in \cite{pietzuch2006network}, and showed that it requires fewer migrations before converging to the optimal placement and fewer control messages. The easiest way to prevent needless migrations is to use a threshold that ensures that they are beneficial. Load balancing systems commonly use thresholds of load imbalance to ensure that the load is redistributed only when the load imbalance is above a threshold. A different type of threshold targets the migration itself to ensure that its benefit is worth its cost. Pietzuch et al. \cite{pietzuch2006network} proposed a method that migrates data only when the benefit in terms of network capacity is higher than a threshold based on the cost of migration.

Using the cost of migration as a goal of optimization means penalizing a placement alternative based on it. Even if a placement is better than the given placement, it might not be preferred because the cost of the reconfiguration is too high. In load balancing-based approaches, the cost of migration is commonly minimized but most often as an implicit goal rather than as part of the objective function. The goal is generally to achieve an acceptable load distribution as quickly as possible, and the redistribution itself constitutes the highest cost. The cost of migration can be minimized while maintaining the load balance \cite{fang2018distributed}. The number of migrations can be minimized while fulfilling QoS requirements \cite{repantis2008hot}. Another way is to maximize the improvement in a query plan and divide the improvement in performance by the cost of migration \cite{lei2014robust}. Gedik et al. \cite{gedik2014partitioning} explored three ways of making load balancing decisions using load and the cost of migration. First, they minimized the cost of migration while the conditions on load balancing were used as constraints. Second, the ideal cost of migration was used as a constraint to minimize load imbalance. Finally, a flexible solution was proposed that combines both load imbalance and the cost of migration as part of an objective function.

In elasticity-based approaches, the cost of migration is often considered in the same way as in load balancing because scaling can be considered to be an extension of load balancing. Zacheilas et al. \cite{zacheilas2015elastic} minimized the monetary costs of computational resources, the cost of migration, and the cost of missing tuples. In this approach, a tradeoff is made between the cost of resources, the cost of missing tuples, and the migration time. A reinforcement learning-based approach was used in \cite{cardellini2018decentralized} that minimizes the cost of reconfiguration, the performance penalty due to QoS constraints, and the cost of resources for using the computational resources.

\subsubsection{Proactive migration decisions}
Current migration solutions generally use reactive approaches to make migration decisions. For instance, a migration might be triggered if a node is overloaded and QoS guarantees are violated, such as when the tuple latency increases excessively. Most proactive solutions predict whether the node can sustain the workload.

Some solutions predict the adaptability of QoS violations \citep{repantis2008hot, lohrmann2015elastic}. Repantis et al. \cite{repantis2008hot} used linear regression and the incoming tuple rate to predict QoS violations of the end-to-end execution time. They predicted QoS violations to prevent them. Lohrmann et al. \cite{lohrmann2015elastic} built a predictive latency model using queuing models to make scaling decisions.

Zacheilas et al. \cite{zacheilas2015elastic} estimated the load and expected latency of Esper to make scaling decisions by using Gaussian processes \cite{rasmussen2003gaussian} because they can help estimate the uncertainty in predictions. However, this method has a cubic computational complexity due to the use of matrix inversion. Wang et al. \cite{wang2017automating} predicted resource usage in real time to choose the configuration that can minimize CPU and memory resources while fulfilling QoS guarantees. This is done using incremental learning techniques based on Weka \cite{weka} and MOA \cite{moa}. De et al. \cite{de2017proactive} used the MPC to predict optimal scaling decisions, called the future horizon. Buddhika et al. \cite{buddhika2017online} used prediction rings to forecast the interference score that expresses the degree to which a system is expected to be overloaded. Lombardi et al. \cite{lombardi2017elastic} used a reactive and a proactive mode for their Elysium system. In the former case, the tuple rate is used as the basis for decisions, and in the latter, the input load is predicted over a certain time, called the prediction horizon. Liu et al. \cite{liu2019adaptive} predicted the load of operators as the number of tuples that operators need to process during a prediction horizon. In WASP \cite{jonathan2020wasp}, the expected input and output rates of the operators are estimated as an alternative to backpressure monitoring for estimating load. Backpressure is weaker as it is based on the observed load instead of the actual workload, and this may lead to less accurate adaptation decisions \cite{kalavri2018three}. A composition of reactive, proactive, and delayed migrations was presented in \cite{lindeberg2022study}. The results of this empirical study indicated that knowledge of the window state can be used to schedule a migration when the state is minimal (i.e., after completing a tumbling window, as in \cite{luthra2018tcep}), or when no output tuple is affected by the migration.

\section{Empirical quantification of core concepts of migration}
\label{sec:experiments}

In this section, we apply the core concepts defined in Section \ref{sec:migration-model} and surveyed in Section \ref{sec:migration-rqs} to gain empirical insights. This is useful for two aspects of migration: the migration algorithm and the decision model for migration. We argue that it is important to model the tradeoff between the cost of migration and its benefit, and that this can be quantified and evaluated. Minor tweaks to the migration algorithm can result in significant changes in its performance and the correctness of its queries, such as an inability to correctly buffering tuples. This highlights the importance of using a common language when defining and using migration algorithms.

We first define two direct moving state migration algorithms: (1) one that uses partial state movement, and (2) another that sends the entire state at once. They are defined in an abstract way such that they can be implemented in the Apache Flink and Siddhi SPEs. We then define decision models to determine when and where to migrate the data. We conducted a real migration experiment to analyze the migration algorithms based on the NEXMark benchmark \cite{tucker2008nexmark}. We also show a use case of the decision models for migration to illustrate their effect on decision-making.

\subsection{Migration algorithms}
The difference between the partial state movement algorithm and the all-at-once state movement algorithm is that the former splits the state into a large static state and a small dynamic state. The static state is transmitted while the operator is still running and processing tuples, followed by the extraction of the dynamic state. As such, static state transmission involves little or no overhead in query processing, and constitutes only one additional step in the algorithm. Note that a partial state movement algorithm might split the state into more than two parts, such as in Megaphone \citep{hoffmann2019megaphone}.

We use the algorithms described in Section \ref{sec:migration-rqs} as basis. In particular, we divide the algorithms into functions that are executed by different nodes participating in the network according to their roles. When moving the state, the old host provides the next hops for the query. Thus, there is no need to add them explicitly in these tasks. These tasks follow a similar format to that used in Expose \cite{volnes2020expose}, which is a framework and toolset for efficiently defining and executing DSPS experiments. Wrappers for different SPEs are provided such that all SPEs support a common set of tasks. Expose has been extended with additional tasks to enable operator migration.

Listings \ref{lst:moving-state-nca} and \ref{lst:moving-state-ca} describe the tasks we use to define the all-at-once state movement algorithm and the partial state movement algorithm, respectively. They differ slightly from similar algorithms in Section \ref{sec:migration-steps-rq} in some respects, such as the ways in which streams are managed. It is possible to send a batch of tasks to upstream nodes, as in Flux \cite{shah2003flux}, to the new host as in \cite{hummer2011dynamic}, and to the old host as in \cite{repantis2007alleviating}. The only difference between the all-at-once state movement and partial state movement algorithms is that the latter involves sending the state of the static query before redirecting the upstream nodes.

\begin{figure}[hbtp]
\vspace{-1em}
\centering
{\begin{minipage}[t]{.44\textwidth}
\begin{flushleft}
\begin{blstlisting}[basicstyle=\scriptsize\ttfamily,caption={All-at-once state movement},label={lst:moving-state-nca}]
ControlMessage(OH 
    ControlMessage(NH,
        RequestMigration(query),
        BufferStreams(Streams(query))
        StopStreams(Streams(query)))
    ControlMessage(Upstream,
        Redirect(Streams(query), OH, NH))
    MoveState(query, NH)
    AddNextHop(Streams(query), NH))
\end{blstlisting}
\end{flushleft}
\end{minipage}
\begin{minipage}[t]{.44\textwidth}
\begin{flushright}
\begin{blstlisting}[basicstyle=\scriptsize\ttfamily,caption={Partial state movement},label={lst:moving-state-ca}]
ControlMessage(OH 
    ControlMessage(NH,
        RequestMigration(query),
        BufferStreams(Streams(query))
        StopStreams(Streams(query)))
    MoveImmutableState(query, NH)
    ControlMessage(Upstream,
        Redirect(Streams(query), OH, NH))
    MoveIncrementalState(query, NH)
    AddNextHop(Streams(query), NH))
\end{blstlisting}
\end{flushright}
\end{minipage}}
\label{fig:moving-state-algorithms}
\vspace{-1em}
\end{figure}

\paragraph{Implementation}
To facilitate the migration of any moving state operator, an SPE needs to be able to extract the runtime state and the load state. This feature is supported in different ways by Siddhi and Flink. In Siddhi, the state is loaded from the runtime system into a byte array, and requires that the entire state is available in memory. As such, there are limitations on how large the state can be. On the contrary, Flink writes the state as a set of checkpoint files, each of which does not exceed a configurable size. Therefore, the state to migrate with Flink can be larger than in Siddhi. The implementation of the other tasks, including {\ttfamily{BufferStreams}}, {\ttfamily{StopStreams}}, {\ttfamily{ControlMessage}}, and {\ttfamily{AddNextHop}}, is supported through simple tasks defined in the SPE wrapper in Expose \cite{volnes2020expose}.

The standard moving state algorithm is implemented in Flink and Siddhi, but only Flink supports partial state movement since this requires the ability to split a given state into a large, immutable state and smaller incremental checkpoints. This feature is supported by one of the state backends in Flink called RocksDB \cite{rocksdb}. Flink with RocksDB is also used for the checkpoint-assisted algorithm in Rhino \cite{del2020rhino}, which uses partial state movement. Another benefit of RocksDB is that it does not store the entire state in memory while the system is running, but instead writes it to file and minimizes its size based on multiple criteria.

\subsection{Decision models}

As discussed in Section \ref{sec:migration-decisions-rq}, there are many different ways of making the migration decision. Our solution is to make the decision process as transparent and meaningful as possible by optimizing the QoS. The goal is to maximize the performance of a placement while penalizing it based on the cost of migration, which varies for different nodes and is zero for the current host. In this way, it is clear why a new placement is selected over the old one, for reasons other than simply that the old host is over-provisioned or the new placement delivers better performance.

The amortization time ($at$) varies depending on the reliability of a placement score for the operator on a given host. If the placement score is stable over time, the amortization time increases since it is less likely that the placement becomes suboptimal shortly after the migration. For instance, a mobile node might have less consistent placement score than a server located in a data center, and as such, it is even more important that the migration is worth the cost of it.

\begin{equation}
at(h, op) = min_{at} + (max_{at}-min_{at})/100 * (100-rsd_{p}(h, op)) 
\end{equation}

where $rsd_{p}(h, op)$ expresses the relative standard deviation (RSD) of the historical placement scores of operator $op$ on host $h$.

When defining the cost of migration, operator downtime alone is not sufficient because it does not reveal how many tuples, if any, are affected by the downtime. Therefore, we use the tuple rate during the migration as a foundation for the cost of migration. Since the data sink waits for tuples from the operator, we consider the number of expected output tuples $PT_{out}(at, op)$ that are affected by the migration to calculate its cost. Buddhika et al. \cite{buddhika2017online} describes a tuple prediction method that can be applied here.

\begin{equation}
PT_{out}(at, op) = PT_{in}(at, op) * Sel(op)
\end{equation}

where $PT_{in}(at, op)$ is the predicted number of input tuples for operator $op$ during amortization time $at$ and $Sel(op)$ is the selectivity of operator $op$.

The cost of migration can be calculated as the operator downtime divided by the amortization time. Since we focus on output tuples from the query, the cost of migration $C(op, oh, nh)$ is defined as the ratio of the predicted output tuples ($PT_{out}(mt(oh, nh, op)$) from a query during migration to the output tuples predicted from it during the amortization time ($PT_{out}(at(nh, op))$).

\begin{equation}
C(op, oh, nh) = w_c * \frac{PT_{out}(mt(oh, nh, op))}{PT_{out}(at(nh, op))}
\end{equation}

The cost has a weight associated with it, meaning that the system can dynamically change how much the cost of migration matters based on the selected policy. If $w_c$ is set to one, this suggests that the performance of a placement should be reduced in proportion to the number of tuples that are received during operator downtime. If $w_c$ is set to 1.5, the placement is penalized further. This makes sense as buffered tuples may take some time to process, during which no new tuples may be processed.

Given the amortization time, the benefit of the migration $B_m(op, oh, nh)$ of a placement is its finite performance penalized by the cost of migration, instead of it being a general placement score. Of two placements with the same migration cost, the one with the higher placement score is selected. The only difference arises when two placements have different costs of migration, for instance, when comparing the given placement with zero cost of migration with another placement that requires a migration.

The benefit of migration can be calculated as
\begin{equation}
B_m(op, oh, nh) = P(nh, op) * (1-C(op, oh, nh))
\end{equation}
where $P(nh, op)$ is the estimated placement score for the new host $nh$ running operator $op$.

The above functions show how the migration decisions are made. Migration checks are periodically performed by calculating the placement score. Following this, the benefit of the migration of placements is calculated by penalizing the placement score based on the cost of migration. We define $M(oh, phs, op)$ as the potential host with the maximum benefit for the given operator. This host is selected as the future host for the operator, and triggers migration if it is not the given placement.

\begin{equation}
M(oh, phs, op) = {\max\limits_{ph \in phs}B(oh, ph, op)}
\end{equation}

\vspace{-1em}
\subsection{Empirical evaluation}

We quantitatively analyzed our proposed decision models for migration through a use case and our migration algorithm through experiments. The goal was to show the usefulness of incorporating the cost of migration into the process. We considered a use case for the decision models because it makes the analysis and discussion of the results easier. On the contrary, implementing and running the migration algorithms on SPEs is necessary to understand the impact of migration.

\begin{figure}[htbp]
\vspace{-3em}
\centering
\begin{subfigure}{.49\textwidth}
\begin{adjustbox}{width=.66\textwidth,center}
\begin{tikzpicture}[->,>=stealth',shorten >=1pt,auto,node distance=3cm,
                    semithick]
  \tikzstyle{every state}=[fill=black!30!green,draw=none,text=white]

  \node[state]           (DDS)               {Join};
  \node[state, draw=none, fill=none]                        (OH) [left of=DDS]  {};
  \node[state,label=below:Data sink]                       (DS) [below of=DDS] {};
  \node[state, draw=none, fill=none]          (NH) [right of=DDS] {};
  \node[state,label=below:Upstream]                      (USA) [above right of=OH] {A};
  \node[state,label=below:Upstream]                      (USB) [above left of=NH] {B};
  
  \node[state, draw=none, fill=none, text=black, align=left] () [above of=DDS] {select P.id\, A.itemName\, A.reserve \\
              from Auction A \\
              join Person P on P.id = A.seller};

  \draw[dotted] (USA) edge node {} (DDS);
  \draw[dotted] (USB) edge node {} (DDS);
  \draw[dotted] (DDS) edge node {} (DS);

\end{tikzpicture}
\end{adjustbox}
\caption{Operator graph}
\label{fig:join-query}
\end{subfigure}
\begin{subfigure}{.49\textwidth}
\begin{adjustbox}{width=.7\textwidth,center}
\begin{tikzpicture}[>=stealth',shorten >=1pt,auto,node distance=3cm,
                    semithick]
  \tikzstyle{every state}=[fill=red,draw=none,text=white]

  \node[state,label=below:Suboptimal host]           (DDS)               {D};
  \node[state,label=below:Old host]                        (OH) [left of=DDS]  {C};
  \node[state,label=below:Data sink]                       (DS) [below of=DDS] {F};
  \node[state,label=below:Optimal host]          (NH) [right of=DDS] {E};
  \node[state,label=below:Upstream]                      (USA) [above right of=OH] {A};
  \node[state,label=below:Upstream]                      (USB) [above left of=NH] {B};

  \path (OH) edge node {}  (DDS);

  \draw[] (DS) edge node {} (OH);
  \draw[] (DDS) edge node {} (NH);
  \draw[] (DS) edge node {} (NH);
  \draw[] (NH) edge node {} (USA);
  \draw (NH) edge node {} (USB);
  \draw (DDS) edge node {} (USB);
  \draw[] (DDS) edge node {} (USA);
  \draw[] (DDS) edge node {} (OH);
  \draw (OH) edge node {} (USA);
  \draw (DDS) edge node {} (DS);
  \draw[] (OH) edge node {} (USB);

\end{tikzpicture}
\end{adjustbox}
\label{fig:topology}
\caption{DSP overlay topology}
\end{subfigure}
\caption{Evaluation scenario}
\label{fig:evaluation-scenario}
\vspace{-1em}
\end{figure}
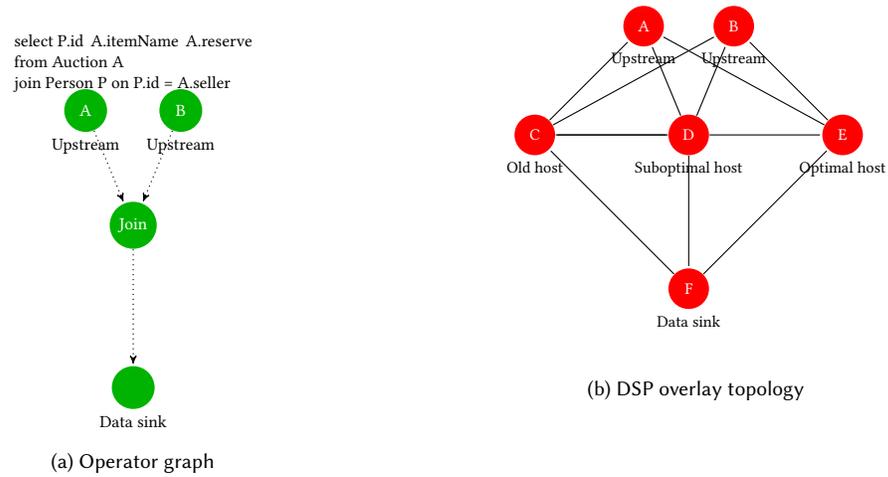

Figure \ref{fig:evaluation-scenario} illustrates our evaluation scenario, Figure \ref{fig:evaluation-scenario}a is the operator graph used for both the use case and the migration experiment, and Figure \ref{fig:evaluation-scenario}b is the DSP overlay topology. The mapping from the operator graph to the physical topology is demonstrated using the decision models in Section \ref{subsubsec:use-case}, and an experiment involving the migration of state from the join operator on one node to another is described in Section \ref{subsubsec:experiment}.

\subsubsection{Decision model use case}
\label{subsubsec:use-case}

The decision models for migration were assessed in this use case. They were applied using a prediction model oracle with 100\% accuracy to make migration decisions. We expect that the migration time can be predicted based on periodically updated topological information and network statistics. By using knowledge of the number of tuples sent in the time window and the migration time, we can predict the total end-to-end latency of the tuples during a given time window. The parameters of the use case are provided in Table \ref{table:use-case-parameters}. We considered two source nodes A and B, three potential hosts C, D, and E, and a sink node F.

\begin{table}[htbp]
\centering
\scriptsize
\begin{tabular}{|c|c|}
\hline
Parameter name & Parameter value \\
\hline
Amortization time & 5 s \\
\hline
Bandwidth C<->D & 200 mbit/s \\
\hline
Bandwidth C<->E & 100 mbit/s \\
\hline
Bandwidth D<->E & 100 mbit/s \\
\hline
Bandwidth Leader<->Hosts & 200 mbit/s \\
\hline
Latency between all links & 1 ms \\
\hline
Control message size & 168 bytes \\
\hline
Migration cost (C) & 0 \\
\hline
Migration cost (D) & 0.1 \\
\hline
Migration cost (E) & 0.5 \\
\hline
\end{tabular}
\caption{Parameters of the use case}
\label{table:use-case-parameters}
\vspace{-3em}
\end{table}

\paragraph{Results}
Table \ref{table:use-case-results} shows the results of the use case for the configurations given in Table \ref{table:use-case-parameters}. The predicted tuples (PT) during the migration kept increasing through the runs, leading to more disruptions to potential migrations. The cost of migrating the nodes was the same in all runs. Node C was the host and, therefore, had no migration time. Node D had a predicted migration time equal to 10\% of the amortization time, and therefore had a cost of migration of 0.1, and Node E had a cost of migration of 0.5 because its estimated migration time was 50\% of the amortization time. The QoS of the host Nodes C and E remained almost constant in all cases, and only that of Node D underwent significant changes. Therefore, the benefit of migration of Node D was also the only one that changed significantly in the different runs. Node E had a better QoS than Node C, at 2.4--2.7 compared with 1.4--1.7. The cost of migration reduced the score of Node E by 50\% because it did not have enough time during the amortization window to pay off the cost of migration. The two right-most columns show the optimal host when the cost model (CM) was considered and when no cost model (NCM) was considered. Node E was the optimal host in all cases in which the cost model was ignored, and was never optimal when the cost model was used because the cost of migration was so high. As Node D incurred a relatively low cost of migration and eventually reached a high QoS, it became the optimal host in Scenarios 3 and 4 when the cost model was used. 

\begin{table}[htbp]
\centering
\footnotesize
\begin{tabular}{lllllllll} \toprule
PT (at) & QoS (C) & $B_m$ (C) & QoS (D) & $B_m$ (D) & QoS (E) & B (E) & P (CM) & P (NCM) \\
1000 & 1.5 & 1.5 & 1 & 0.85 & 2.7 & 1.35 & C & E \\
2000 & 1.6 & 1.6 & 1.6 & 1.36 & 2.5 & 1.25 & C & E \\
3000 & 1.4 & 1.4 & 2.5 & 2.125 & 2.4 & 1.2 & D & E \\
4000 & 1.7 & 1.7 & 2.8 & 2.38 & 2.6 & 1.3 & D & E \\
\bottomrule
\end{tabular}
\caption{Results of the use case }
\label{table:use-case-results}
\vspace{-2em}
\end{table}

Because the QoS was stable for Node E, and was significantly better than that for Node C, it was possible to dynamically increase the amortization time for nodes that had demonstrated their stability in terms of the predicted QoS. Node D, on the contrary, has a significantly variable QoS, which increased above that of Node C with a value of 1.6 in the second row, but yielded a lower benefit of migration of 1.36, and was not selected as the new host. With a score of 2.5 that was reduced to 2.125 given the migration cost, it beat the given host, and was selected as the new host.

\subsubsection{Migration experiment}
\label{subsubsec:experiment}
In this experiment, we evaluated the proposed migration algorithm by analyzing its execution and comparing its results with those obtained when executed with two SPEs. The all-at-once state movement runs were used to send 100,000, 1,000,000, and 5,000,000 tuples. The partial state movement runs were used to send 1,100,000 and 5,100,000 tuples. The additional 100,000 tuples with partial state movement were sent during migration, and were part of the dynamic state to be sent. The experiment used a simplified version of the topology in Figure \ref{fig:evaluation-scenario}b in two ways. First, there was only one upstream node. Second, there were only two hosts: the old and the new host.

The experiment tested the cost of migration by varying the size of the state to be moved. For runs of the partial state movement, the number of tuples that were migrated during static state migration and dynamic state migration were varied. The dataset of the NEXMark stream processing benchmark \cite{tucker2008nexmark} was used in the experiment. NEXMark is based on an auction scenario, where three streams are used: a person, a bid, and an auction item stream. For this experiment, only one of the queries was used, one that joined the person and bid streams. We used this query because a join query makes it easier to test the migration algorithm and adjust the size of the state to migrate. One can simply send a given number of tuples of the first stream, migrate it to the new host, and send a single tuple of the second stream to the new host. If this triggered the correct number of output tuples to be produced, the migration was considered to have been successful.

Four processes with different roles were used in the experiment: a data producer node, a host running the operator to be migrated, a new host that contained the operator after migration, and the data sink that consumed the output tuples of the operator. We used two machines for the experiment, one for the old host, and the other to run the data producer, data consumer, and new host. The specifications of the machines are shown in Table \ref{table:specifications}. In the experiment, the data producer generated a certain amount of auction tuples that were sent to the old host. The state was then migrated to the new host, and the data producer sent a single person tuple that joined with all the auction tuples to trigger the same number of output tuples to be sent to the data sink as auction tuples that were sent prior to the migration. The query we used was a modification of NEXMark's \cite{tucker2008nexmark} Query 8. Originally, this query does not select the itemName of the auction, but chooses the person's name. Each auction tuple was augmented with 1 kB of a randomized string to increase the size of the state to be migrated.

\begin{table}[htbp]
\centering
\scriptsize{
\begin{tabular}{|c|c|c|c|}
\hline
OS & CPU & RAM & Description \\
\hline
Ubuntu 20.04.2 & Intel Xeon Gold 5215 2.50 GHz & 45.8 GB  & Old host \\
\hline
Ubuntu 18.04.4 & Intel Core i7-7800X 3.50 GHz & 62.5 GB & New host, data producer, data sink, Expose coordinator \\
\hline
\end{tabular}
\caption{Server specification}
\label{table:specifications}
}
\vspace{-2em}
\end{table}

In all runs, we counted the number of tuples that were migrated, the state size, the state extraction time, the state transfer time, and the state loading time. For the partial state movement algorithm, the same parameters were used for the static and dynamic states. The state to be migrated was ranged from 1 to 5 GB; however, Siddhi has a limit of 1 GB because it extracts the entire state into a single byte array, whereas Flink's state backend RocksDB splits the state into multiple files.

\paragraph{Results}
Tables \ref{table:nca-experiment-results} and \ref{table:ca-experiment-results} show the results of the experiments, where the former shows the outcomes of the all-at-once state movement algorithm and the latter those of the partial state movement algorithm. Siddhi and Flink migrated operator states of different sizes depending on the query and the number of tuples that were processed.

\begin{table}[htbp]
\centering
\footnotesize
\begin{tabular}{lllllll} \toprule

SPE & \# Tuples & State size & State extraction & State loading & State transfer  & Freeze time \\

Siddhi & 100,000 & 100 MB & 2.8 s & 1.76 s & 0.94 s & 3.6 s \\

Siddhi & 1,000,000 & 1 GB & 21.3 s & 13.6 s & 9 s & 43.9 s \\

Flink & 100,000 & 100 MB & 270 ms & 2.267 s & 1.08 s & 3.6 s \\

Flink & 1,000,000 & 1 GB & 3.329 s & 20.295 s & 11.582 s & 35.2 s \\

Flink & 5,000,000 & 5 GB & 19.781 s & 77.411 s & 52.759 s & 150 s \\
\bottomrule
\end{tabular}
\caption{Results of all-at-once moving state experiment }
\label{table:nca-experiment-results}

\resizebox{1\textwidth}{!}{
\begin{tabular}{lllllllllll} \toprule

SPE & Tuple count (S) & Tuple count (D) & Size (S) & Size (D) & State extraction (S) & State extraction (D) & State loading & Transfer (S) & Transfer (D) & Freeze time \\

Flink & 1,000,000 & 100,000 & 1 GB & 113 MB & 76 ms & 149 ms & 1.77 s & 11.639 s & 111 ms & 2 s \\

Flink & 5,000,000 & 100,000 & 5.2 GB & 376 MB & 363 ms & 528 ms & 3.031 s & 57.186 s & 4.295 s & 7.8 s \\
\bottomrule
\end{tabular}}
\caption{Partial moving state experiment results}
\label{table:ca-experiment-results}
\vspace{-2em}
\end{table}

The state transfer times of Siddhi and Flink were similar because they used similar implementations of the TCP socket. Siddhi performed slightly better, where this can be explained by the fact that Flink read the checkpoint from multiple files, and state transfer in it was executed in parallel with reading the files. State extraction appeared to scale relatively poorly for both Siddhi and Flink with the all-at-once state movement algorithm, but with the partial moving state, Flink had a significantly lower state extraction overhead. Moreover, state loading using partial state movement was much faster than without it. Note that these results do not represent the general performance of the SPEs, but the outcomes for a specific join query that was used for a specific systems. Another query might have yielded different results. For instance, this query was very write heavy. All the received auction tuples were used in the migration, and only once they had been read, i.e., after the person tuple had been received on the new host. In this case, the partial state movement algorithm performed better in all respects.

One might think that the all-at-once state movement algorithm would have had faster state loading as it has a monolithic checkpoint, but this was not the case. We think this result is obtained because the incremental checkpointing uses RocksDB’s native checkpoint files whereas Flink's full snapshot approach iterates through the RocksDB state and creates its own files. RocksDB is designed to be efficient, and performs indexing to increase its efficiency. This benefit was lost in the full snapshot approach.

If we assume that the number of tuples that were received during the freeze time arrived at a fixed rate, the average additional tuple latency as a result of the migration was equal to half the freeze time. The maximum additional tuple latency was approximately equal to the freeze time and the minimum was close to zero. The number of affected tuples could vary significantly, ranging from zero to hundreds of thousands per second.

The partial state movement algorithm performed much better than the all-at-once algorithm in terms of freeze time, even without counting the poor state loading performance of the all-at-once algorithm. The reason is that most of the state was moved before the operator was shut down. This difference in performance was especially significant when considering how similar the algorithms were in terms of how they were described in Listings \ref{lst:moving-state-nca} and \ref{lst:moving-state-ca}. Only one task was added to \ref{lst:moving-state-ca}, which was to migrate the immutable state before the streams were redirected by the upstream nodes. This leads to the important conclusion that the literature can benefit from a common language when defining or using a migration algorithm. Exactly what tasks are executed during the migration, in particular, those that increase the freeze time, can be described using, e.g., the concepts described in the migration model in Section \ref{sec:migration-model}.

The proposed migration model can foster the development of new migration algorithms, and can help avoid duplicate solutions. One explanation for why multiple, similar migration algorithms have been developed is that the algorithm is typically only part of the contribution of approaches, along with decision models for migration, the system itself, and its evaluation.

\section{Reflections and Future Directions}
\label{sec:reflections}
The historical development in data migration, from the early single-track moving all-at-once state migration solutions to checkpoint-assisted partial state movement and parallel-track solutions without state movement, has been driven by the deployment of SPEs to the cloud environment, and improvements to them to achieve fault tolerance and dynamic scalability. Cloud environments provide large amounts of computational resources (even though at different scales), and their servers are interconnected with low-latency high-bandwidth networks. Therefore, advanced state management solutions in cloud-based systems might work well in fog environments.

However, in fog environments that are geo-distributed, the connections between hosts have substantially lower available bandwidth and higher latencies that can impact the cost and benefit of operator migration, and require adapted migration mechanisms. The periodic checkpointing and replication of checkpoints are used in some cluster-based SPEs to facilitate fault tolerance and fast migrations, but it is not always feasible to replicate and distribute checkpoints, especially in resource-constrained IoT devices. For future in-network processing solutions with mobile platforms, e.g., advanced crowd-sensing applications, energy is an important factor to consider for operator placement, the design of migration mechanisms, and the calculation of cost and benefit. This survey shows that energy is not yet an issue for state-of-the-art operator migration approaches.

It is clear that the smaller the size of the data to be migrated is, the less is the energy that is consumed. Therefore, scheduling operator migration at a point in time when the state is small or even zero is important. This can be achieved, for example, by delayed migration by waiting until a tumbling window is emptied \cite{ottenwalder2014mcep}, and through proactive migration. Another alternative is to allow for some inconsistent state where not all of it is migrated to the new host. In some cases, aggregation operators can be moved without the state, resulting in zero freeze time. Alternatively, load shedding techniques can be applied to send some of the state, or components of it can be assigned a priority such that only the most important state is migrated, while the less significant part of it is omitted. However, a thorough investigation of the pros and cons of reactive, delayed, and proactive migrations in different environments with different workloads and guarantees of consistency is still elusive. 

Another gap in research is an analysis and comparison of stream management techniques. Several aspects are important for such an investigation: (1) the sequence of tasks like the stopping, buffering, redirecting, and starting of streams, (2) the locations where streams are buffered, (3) the delivery semantics, i.e., at least once, at most once, exactly once as well as ordered or out-of-order delivery, (4) and tasks related to buffer management and transport protocols. 

The quality of decision-making on migration depends on the data available to calculate its cost and benefit, as well as the freshness of the data. The continuous collection and dissemination of monitoring data in a DSPS can be expensive. Efficient monitoring solutions, and leveraging other sources of monitoring data that are, for example, used for network and system management have the potential to reduce the overall cost of a DSPS and ensure good decision-making. 

Leveraging historical data to perform predictions with advanced statistics or modern machine learning solutions, as is done for traffic prediction in network management \cite{abbasi2021deep} and data prediction in wireless sensor networks \cite{dias2016survey}, is another subject that deserves attention in research. Both proactive migration and the use of amortization time in the cost model require some form of prediction. The oxymoron of operator migration, i.e., that the need for migration occurs when the cost of migration is high, can be avoided with proactive migration. Furthermore, proactive migration can be used to schedule a migration when the state is still small in size. However, both traffic and data patterns might be changing during the deployment of DSPSs, and appropriate and efficient online learning solutions need to be investigated for operator migration.

\section{Conclusions}
\label{sec:conclusion}
DSP is becoming increasingly important for handling data with high velocity and a large variety. The variety of data from different sources and over time as well as other system dynamics, e.g., resource availability, require adapting distributed stream processing accordingly. Operator migration is the mechanism for keeping a DSPS in an "optimal" configuration over its lifetime. This survey provided an overview of solutions for operator migration from a historical perspective and that of the goal of migration. Both perspectives show that the deployment environments and the purpose of the system have a strong impact on the design of migration mechanisms. Unfortunately, the terminology in this area is not always consistent. Therefore, we introduced a conceptual model of operator migration based on the largest common denominator in the literature to introduce a common/unified terminology. The model facilitated the classification of existing solutions and structured their description with respect to two research questions: (1) Which mechanisms are used to perform migration? (2) How is the migration decision executed? Emphasis was placed on its costs and benefits. These aspects are important for operator migration, but are often only implicitly addressed or are neglected altogether. The description of existing solutions shall provide the reader with a good understanding of the design alternatives from an algorithmic viewpoint. We complemented this with an empirical study to give the reader some quantitative insights into the impact of different design alternatives for migration mechanisms (i.e., all-at-once and partial state movements), and the impact of the choice of data stream processing (i.e., Siddhi and Apache Flink).

\section{Acknowledgments}
This work was supported by the Parrot Project (Research Council of Norway, IKTPluss, 311197).
The authors thank Fabrice Starks and Stein Kristiansen for insightful discussions on mechanisms of operator migration, and Boris Koldehofe for reviewing and commenting on an earlier version of the manuscript.

\bibliographystyle{unsrt}
\bibliography{bibliography}

\end{document}